\documentclass[12pt]{article}

\usepackage[T1]{fontenc}%
\usepackage{graphicx}%
\usepackage{longtable}%
\usepackage{wrapfig}%
\usepackage{rotating}%
\usepackage[normalem]{ulem}%

\usepackage{amsmath}%
\usepackage{amssymb}%
\usepackage{authblk}%
\usepackage{capt-of}%
\usepackage{enumitem}%
\usepackage[margin=1in]{geometry}%
\usepackage{hyperref}%
\usepackage{import}%
\usepackage{mathrsfs}%
\usepackage{physics}%
\usepackage{subcaption}%
\usepackage{tensor}%
\usepackage{titlesec}%
\usepackage{titling}%
\usepackage{xspace}%
\usepackage{xparse}%

\usepackage{cleveref}

\newcommand{\cosm}{\mathrm{c}}%
\newcommand{\ent}{\mathrm{e}}%
\newcommand{\New}{\mathrm{N}}%
\newcommand{\pln}{\mathbb{C}}%
\newcommand{\tor}{\mathrm{t}}%
\newcommand{\uhp}{\mathbb{C}/\mathbb{Z}_2}

\newcommand{\eg}{\textit{e.g.\@\xspace}}%
\newcommand{\ie}{\textit{i.e.\@\xspace}}%
\newcommand{\via}{\textit{via.\@\xspace}}%

\author{Hong Zhe Chen}%
\affil{
  \href{mailto:hzchen@ucsb.edu}{hzchen@ucsb.edu} \\
  \\
  Department of Physics, University of California,\\
  Santa Barbara, CA 93106, USA}%
\date{}%
\title{Disentanglement as a strong cosmic censor}%

\hypersetup{%
  pdfauthor={Hong Zhe (Vincent) Chen},%
  pdftitle={Disentanglement as a strong cosmic censor},%
  pdfkeywords={},%
  pdfsubject={},%
  pdflang={English}%
}

\begin{document}

\begin{titlingpage}
  \maketitle

  \begin{abstract}
    If entanglement builds spacetime, then conversely, disentanglement ought to
    destroy spacetime. From the quantum null energy condition and quantum
    focusing conjecture, we obtain disentanglement criteria which necessitate
    infinite energies and strong spacetime singularities. We apply our results
    to the strong cosmic censorship proposal, where singularities at the Cauchy
    horizons in black holes are desirable. Using our disentanglement criteria
    and without resorting to any detailed calculations, we provide an
    exceedingly general and physically transparent discussion of strong cosmic
    censorship in semiclassical black holes. We argue that strong cosmic
    censorship is enforced in asymptotically flat and de Sitter black holes by
    disentanglement and describe how similar disentanglement might be avoided in
    some anti-de Sitter cases.
  \end{abstract}
  \vfill
  \noindent This essay received an Honorable Mention in the 2024 Essay
  Competition of the Gravity Research Foundation. Extra commentary and examples
  have been added as footnotes and an appendix.
\end{titlingpage}

\tableofcontents



\section{Introduction}

A motif that appears in the study of quantum information in gravity is the
notion that entanglement builds spacetime geometry \cite{VanRaamsdonk:2010pw}.
Taken literally, the entanglement between naively distant degrees of freedom is
recast as geometric wormholes \cite{Maldacena:2013xja}. More abstractly, the
identification of black hole interiors with Hawking radiation through
entanglement has recently revealed a partial resolution
\cite{Almheiri:2019psf,Penington:2019npb} to the infamous information paradox
\cite{Hawking:1976ra,Almheiri:2012rt}. In this essay, we will explore the
converse idea: disentanglement breaks apart spacetime geometry
\cite{VanRaamsdonk:2010pw,Czech:2012be,Emparan:2023ypa}.

We will focus particularly on short-distance entanglement in quantum field
theories (QFTs) on semiclassical backgrounds, and the consequences of destroying
this entanglement. To connect entanglement to energy and its imprint on
spacetime geometry, we will appeal to the quantum null energy condition (QNEC)
and quantum focusing conjecture (QFC) \cite{Bousso:2015mna,Bousso:2015wca}. As
successors of the classical null energy condition (NEC) and focusing theorem,
the QNEC and QFC incorporate quantum information into bounds that govern QFTs in
semiclassical spacetimes. These considerations will motivate disentanglement
criteria which unavoidably lead to infinite energies and strong spacetime
singularities.

Such singularities are sometimes desirable. In particular, certain black holes
contain Cauchy horizons which delimit the unique evolution from initial data or
states on an initial time slice. Naively, the breakdown of the theory's
predictive power beyond the Cauchy horizon seems problematic. However, strong
cosmic censorship (SCC) postulates that solutions will generically develop
singularities at or before the Cauchy horizon and cannot be further extended
\cite{Penrose:1974cup}. Though classically violated in certain situations, SCC
seems to be restored semiclassically by quantum fields \cite{Hollands:2019whz}.

Using our disentanglement criteria, we will provide an exceedingly general
argument for SCC that can be narrated by words and pictures alone. Our story
cohesively ties together results previously reliant on explicit calculations set
in certain backgrounds
\cite{Zilberman:2019buh,Hollands:2019whz,Hollands:2020qpe,Zilberman:2022aum}, in
particular generalizing and strengthening previous considerations of
entanglement in the context of SCC \cite{Papadodimas:2019msp}. With this
application to SCC, we aim to illustrate the useful sensitivity of spacetime
geometry to the destruction of entanglement.

\section{Entanglement at short distances}

The entanglement of interest to us will be between spatial subregions --- let us
begin by quantifying this entanglement. In QFT, the state of quantum fields in a
spatial subregion \(R\) can be formally described by a density matrix
\(\rho_R\). For example, if \(R\) is a subregion of a full time slice \(R\cup
\bar{R}\) on which the quantum fields are in a pure state \(\ket{\psi}\), then
the reduced state \(\rho_R = \tr_{\bar{R}} \ket{\psi}\bra{\psi}\) on \(R\) would
be obtained by tracing over the Hilbert space in the complementary subregion
\(\bar{R}\). A simple measure of entanglement between \(R\) and its complement
is then given by the von Neumann entropy,
\begin{align}
  S_R
  &= -\log(\rho_R \log \rho_R)
    \;.
    \label{eq:vnentropy}
\end{align}
For our purposes, it will be helpful to isolate the entanglement strictly
between two subregions \(R_1\) and \(R_2\) which need not be complementary. The
relevant quantity in this regard is the mutual information
\begin{align}
  I_{R_1:R_2}
  &= S_{R_1} + S_{R_2} - S_{R_1\cup R_2}
    \;,
    \label{eq:mi}
\end{align}
which subtracts off the contribution from entanglement with everything outside
\(R_1 \cup R_2\).

Both the entropy \labelcref{eq:vnentropy} and the mutual information
\labelcref{eq:mi} between \emph{adjacent} subregions are positively divergent in
QFT, due to the entanglement between degrees of freedom supported arbitrarily
close to the entangling surfaces \(\partial R\) and \(\partial R_1 \cap \partial
R_2\). In smooth states, the excitations of these degrees of freedom are
exponentially suppressed in energy above the locally Minkowskian vacuum.
Nonetheless, there are significant correlations in the quantum fluctuations
across the entangling surface, as visible, for example, from a locally Rindler
perspective \cite{Unruh:1976db,Papadodimas:2019msp}.\footnote{Locally, one can
  view null hypersurfaces orthogonal to the entangling surface as Rindler
  horizons orthogonal to a bifurcation surface in an approximate Minkowski
  spacetime. Ref.~\cite{Papadodimas:2019msp} showed that smooth states of free
  fields have localized Rindler-like modes which are thermally entangled across
  each null surface, much like in Minkowski spacetime \cite{Unruh:1976db}. The
  Boltzmann factor controlling this thermal entanglement suppresses only high
  Rindler-frequency modes. However, for a fixed Rindler frequency, there are a
  logarithmically infinite number of modes supported arbitrarily close to the
  entangling surface. (Of course, their small support implies large Minkowskian
  energy and momentum orthogonal to the entangling surface.) Thus, these modes
  make at least a logarithmically infinite contribution to entanglement entropy
  and the mutual information of adjacent subregions. One expects there also to
  be infinite contributions from other modes with large momenta in directions
  along the entangling surface. \label{foot:rindler}} Fortunately, the same
short-distance entanglement structure is shared by all smooth states. The
divergences in \cref{eq:vnentropy,eq:mi} can therefore be cancelled by
state-independent counterterms, leaving renormalized entropies and mutual
information which are finite in smooth states
\cite{Susskind:1994sm}\footnote{The appendix of ref.~\cite{Bousso:2015mna}
  provides a review.}. From here on, we will be referring to these renormalized
quantities.

A crucial point is that renormalized entropy \(S\) and mutual information \(I\)
can be made arbitrarily negative\footnote{Having a negative renormalized entropy
  or mutual information between adjacent subregions often indicates that one has
  begun probing (in this case, disentangling) degrees of freedom shorter than
  the renormalization distance scale. One may wish instead to set a
  renormalization distance scale shorter than all degrees of freedom under
  consideration; in this case, the renormalization scale is similar to a cutoff
  scale and, as we consider ever shorter distances, renormalized quantities,
  such as entropy in smooth states, become large similar to bare quantities.
  What will be important to us is only that entropy and mutual information for
  adjacent subregions can be decreased by an arbitrarily large amount relative
  to smooth states by disentangling degrees of freedom at arbitrarily short
  distances. For succinctness, we will fix a finite renormalization scale and
  say that renormalized entropy and mutual information become arbitrarily
  negative. (This is, of course, up until the field theory description breaks
  down, such as at the Planck scale. See \cref{foot:planck}.)} by disentangling
short-distance degrees of freedom across the entangling surfaces. What are the
consequences of doing so? Our purpose will be to answer this question in terms
of energy and its backreaction on spacetime geometry. Serving as our bridge from
quantum information will be the QNEC and QFC
\cite{Bousso:2015wca,Bousso:2015mna}, to which we now turn.

\section{The QNEC and the QFC}
\label{sec:qnecqfc}

\begin{figure}
  \begin{subfigure}[b]{0.4\textwidth}
    \begin{center}
      \includegraphics[scale=1.25]{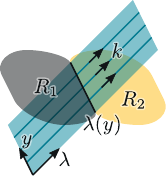}
    \end{center}
  \end{subfigure}
  \hfill
  \begin{subfigure}[b]{0.4\textwidth}
    \begin{center}
      \includegraphics[scale=1.25]{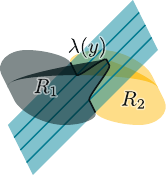}
    \end{center}
  \end{subfigure}
  \caption{\label{fig:qnecqfc_mi}Deformation of the entangling surface
    separating \(R_1\) and \(R_2\) along a null congruence.}
\end{figure}

Though originally formulated in terms of entropy, let us instead state the QNEC
and QFC in terms of mutual information. As illustrated in \cref{fig:qnecqfc_mi},
we consider two spatial subregions \(R_1\) and \(R_2\) sharing a portion of
their boundaries, which we refer to as the entangling surface. From the
entangling surface, an orthogonal congruence of null geodesics is generated by a
null affine vector field \(k\). Let us call the corresponding affine parameter
along the null lines \(\lambda\) and coordinatize spatial directions in the null
congruence by \(y^i\). The entangling surface is a spatial section of the
congruence specified by a function \(\lambda(y)\) and has an induced metric
\(h_{ij}(y,\lambda(y))\). We can then consider deforming \(R_1\) and \(R_2\)
such that the entangling surface \(\lambda(y)\) moves along the null congruence.
In this way, the mutual information \(I_{R_1:R_2}\) can be viewed as a
functional \(I[\lambda]\), whose variation appears in the QNEC and QFC.

In particular, the QNEC \cite{Bousso:2015mna,Bousso:2015wca} states that the
null component of the quantum field stress tensor expectation value is bounded
from below,
\begin{align}
  \ev{T_{kk}(y,\lambda(y))}
  &\ge \frac{1}{4\pi \sqrt{h(y,\lambda(y))}}
    \fdv[2]{I[\lambda]}{\lambda(y)}
    \;,
    \label{eq:qnec_mi}
\end{align}
by the diagonal piece of the second variation,
\begin{align}
  \fdv{\lambda(y_2)}\fdv{I}{\lambda(y_1)}
  &=
    \fdv[2]{I}{\lambda(y_1)}
    \delta(y_1 - y_2)
    + \left(
    \fdv{\lambda(y_2)}\fdv{I}{\lambda(y_1)}
    \right)_{\text{off-diag}}
    \;.
    \label{eq:diagdef}
\end{align}
(The off-diagonal piece is non-positive, simply by the strong subadditivity
property of entropy \cite{Bousso:2015mna}.) The classical limit of QNEC is the
classical NEC \(T_{kk}\ge0\) which leads to the focusing theorem
\(\partial_\lambda \theta \le 0\) governing the classical expansion
\begin{align}
  \theta(y,\lambda)
  &= \frac{1}{\sqrt{h(y,\lambda)}} \pdv{\sqrt{h(y,\lambda)}}{\lambda}
    = \frac{1}{\sqrt{h(y,\lambda)}} \fdv{A[\lambda]}{\lambda(y)}
    \;,
\end{align}
in the area \(A\) of the spatial sections of the null congruence. Similarly
associated to the QNEC is the QFC \cite{Bousso:2015mna} governing a quantum
expansion,
\begin{align}
  \fdv{\bar{\Theta}[\lambda](y_1)}{\lambda(y_2)}
  \le& 0\;,
  &
    \bar{\Theta}[\lambda](y)
  &= \theta(y,\lambda(y))
    + \frac{2G_\New}{\sqrt{h(y,\lambda(y))}}
    \fdv{I[\lambda]}{\lambda(y)}
    \;.
    \label{eq:qfc_mi}
\end{align}

The QNEC and QFC have provided invaluable connections between quantum
information and the more familiar notions of energy and geometry encountered in
semiclassical gravity, notably in black hole and cosmological contexts
\cite{Bousso:2015mna,Bousso:2015eda,Almheiri:2019yqk,Hartman:2020khs}. When
gravity is holographically dual to a theory on its boundary, the QNEC and QFC in
gravity are inextricably tied to consistency conditions in the boundary theory
\cite{Akers:2016ugt,Akers:2017ttv}. Moreover, as a statement purely in QFT, the
QNEC has been independently proven in a variety of settings
\cite{Wall:2011kb,Bousso:2015wca,Balakrishnan:2017bjg}.

Below, we will apply the QNEC and QFC to situations where significant
entanglement is broken between degrees of freedom at short distances. A minor
caveat is that the QNEC \labelcref{eq:qnec_mi} is only expected to hold at
points of the null congruence with vanishing expansion \(\theta\) and shear
\cite{Bousso:2015mna,Akers:2017ttv}. However, we will be more generally
extrapolating the lessons we learn from it at short-distances, where background
curvature should not significantly alter our qualitative results. In contrast,
the QFC and our arguments based on it have no such caveat.

\section{Disentanglement and its consequences}
\label{sec:disentanglement}
Using the QNEC and QFC, we now argue that infinite energies and strong spacetime
singularities become inevitable once certain disentanglement criteria are met.
In particular, we study situations where degrees of freedom in \(R_1\) and
\(R_2\) become increasingly disentangled as the entangling surface is pushed
towards a given spatial section \(\lambda_0(y)\) of the previously introduced
null congruence. Considering a one parameter family of entangling surfaces
\(\lambda_\alpha(y)\) moving monotonically \(\partial_\alpha \lambda_\alpha \ge
0\), our disentanglement criteria will dictate the bad behaviour of mutual
information \(I[\lambda_\alpha]\) as \(\alpha\to 0^-\).

More precisely, it follows from the QNEC \labelcref{eq:qnec_mi} that the
following three conditions cannot simultaneously hold\footnote{In this essay, we
  are studying entanglement at very short distances. Going to ever short
  distances, eventually, the field theory description of physics will break
  down, \eg{} due to quantum gravity near the Planck scale. In this essay,
  writing ``\(\infty\)'' will often come with this caveat. \label{foot:planck}}:
\begin{enumerate}[nosep,label=(\roman*)]
\item{\label{item:disentanglement1} the disentanglement criterion
    \(\eval{\partial_\alpha I[\lambda_\alpha]}_{\alpha\to 0^-} = -\infty\),}
\item{\label{item:smoothnessafter1} regularity \(\eval{\partial_{\alpha}
      I[\lambda_\alpha]}_{\alpha'} > -\infty\) at some \(\alpha'>0\), and}
\item{\label{item:finiteenergy} the finiteness of null energy
    \(\int_{0^-}^{\alpha'} \dd{\alpha}\int\dd{y} (\partial_\alpha
    \lambda_\alpha)^2 \sqrt{h} \ev{T_{kk}} < +\infty\) in the region bounded by
    \(\lambda_0(y)\) and \(\lambda_{\alpha'}(y)\).}
\end{enumerate}
Thus, if we assume that the spacetime and state has an extension beyond the
disentangling surface \(\lambda_0\) that is sensible --- in particular,
satisfying \labelcref{item:smoothnessafter1} --- then there must be infinite
energy present. Of course, this is a contradiction if ``sensible'' also means
having finite energy.

From the QFC \labelcref{eq:qfc_mi}, we can deduce an analogous set of
inconsistent conditions. They are most cleanly stated by choosing the family of
entangling surfaces such that the combination \(\sqrt{h(y,\lambda_\alpha(y))}
\partial_\alpha \lambda_\alpha(y)\) is independent of \(\alpha\). Then, we find
that the disentanglement criterion \labelcref{item:disentanglement1} also cannot
hold simultaneously with
\begin{enumerate}[resume*]
\item{\label{item:smoothnessafter2}regularity \(\eval{\partial_{\alpha} (A+2
      G_\New I)[\lambda_\alpha]}_{\alpha'} > -\infty \) at some \(\alpha'>0\),
    and}
\item{\label{item:finiteexpansion}finite expansion \(\eval{\partial_\alpha
      A[\lambda_\alpha]}_{\alpha \to 0^-} < +\infty\) at \(\lambda_0\).}
\end{enumerate}
Moreover, if, instead of statement \labelcref{item:disentanglement1}, one has
\begin{enumerate}[nosep,label=(\roman*'),start=1]
\item{\label{item:disentanglement3} the stronger disentanglement criterion
    \(I[\lambda_{0^-}]-I[\lambda_{\alpha''}] = -\infty\) for some
    \(\alpha''<0\),}
\end{enumerate}
then statement \labelcref{item:finiteexpansion} can likewise be replaced by the
requirement of
\begin{enumerate}[nosep,label=(\roman*'),start=5]
\item{\label{item:finitearea} finite area
    \(A[\lambda_{0^-}]-A[\lambda_{\alpha''}] < +\infty\).}
\end{enumerate}
In any case, the interpretation parallels that of statements
\labelcref{item:disentanglement1,item:smoothnessafter1,item:finiteenergy}.
Initially supposing an extension beyond the disentangling surface \(\lambda_0\)
which is sensible --- in particular, satisfying statement
\labelcref{item:smoothnessafter2} --- we are invariably led to the contradictory
conclusion that a geometric singularity lives at \(\lambda_0\). This singularity
is so strong that the putative extension of the spacetime cannot solve the
gravitational equations of motion even in a weak sense.

We have thus made concrete the notion that sufficient disentanglement leads to
infinite energies and strong singularities in backreacted spacetimes. There are
many examples, known for some time, of states with infinite energies associated
to disentanglement. These include Rindler vacua
\cite{Parentani:1993yz,Czech:2012be} and states produced by joining quenches of
initially disentangled subsystems \cite{Calabrese:2007mtj, Ugajin:2013xxa,
  Shimaji:2018czt}.\footnote{In \cref{sec:joiningquench}, we review the example
  of a joining quench for two-dimensional CFTs. We show that our disentanglement
  criterion \labelcref{item:disentanglement1} (and even the stronger criterion
  \labelcref{item:disentanglement3}, though gravity is not dynamical in this
  example) is satisfied on a horizon emanating from the joint. Living on this
  horizon is also a null energy shock wave that takes the form of a positive,
  non-integrable contact term.} The latter are closely related to even earlier
work on Lorentzian backgrounds undergoing topology change
\cite{Anderson:1986ww}, where the resulting infinite energies were argued to
violently abort such processes. In similar spirit, we will apply the
disentanglement criteria identified above to argue that strong singularities
develop at black hole Cauchy horizons, effectively aborting the spacetime just
short of the unpredictable region beyond.

\section{Strong cosmic censorship}
\label{sec:scc}

\begin{figure}
  \begin{subfigure}[t]{0.3\textwidth}
    \begin{center}
      \includegraphics[scale=1.25]{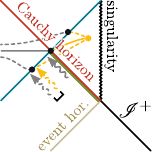}
    \end{center}
    \caption{\label{fig:cauchyhorizon_flat}Asymptotically flat.}
  \end{subfigure}
  \hfill
  \begin{subfigure}[t]{0.3\textwidth}
    \begin{center}
      \includegraphics[scale=1.25]{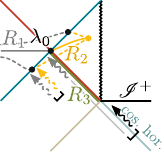}
    \end{center}
    \caption{\label{fig:cauchyhorizon_ds}Asymptotically dS.}
  \end{subfigure}
  \hfill
  \begin{subfigure}[t]{0.3\textwidth}
    \begin{center}
      \includegraphics[scale=1.25]{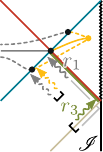}
    \end{center}
    \caption{\label{fig:cauchyhorizon_ads}Asymptotically AdS.}
  \end{subfigure}
  \caption{\label{fig:cauchyhorizon}Black holes with Cauchy horizons.}
\end{figure}

Let us consider semiclassical theories of matter and gravity on black hole
backgrounds which possess Cauchy horizons. Some examples are illustrated in
\cref{fig:cauchyhorizon}, but we expect our discussion to generalize easily to
rotating black holes and potentially to evaporating black holes. Naively, it may
appear that the background and quantum fields can be extended beyond the Cauchy
horizon, but this would be problematic as such extensions are not uniquely
determined by the initial data or state on a full time slice in the far past.
The strong cosmic censorship (SCC) proposal instead suggests that singularities
must appear at (or before) the Cauchy horizon with sufficient strength to cut
off the spacetime above.

Over half a century ago, Penrose \cite{Penrose:1968ar} first argued classically
for SCC by appealing to the arbitrarily strong blue-shift affecting
perturbations falling into the black hole at late times. This is thought to
result in signals carrying arbitrarily large energy, illustrated by the first
wavy arrow \(r_1\) immediately below each Cauchy horizon in
\cref{fig:cauchyhorizon}, which sources a spacetime singularity enforcing SCC.
Unfortunately, counterexamples exist in asymptotically de Sitter (dS)
spacetimes. Here, a competing cosmological red-shift can weaken the
aforementioned effect such that the spacetime may admit an extension beyond the
Cauchy horizon \cite{Hintz:2015jkj,Cardoso:2017soq}. Fortunately, at the
semiclassical level, the quantum stress tensor diverges sufficiently strongly to
restore SCC in certain situations such as Reissner-Nordstr\"om-dS, as verified
by direct calculation \cite{Hollands:2019whz}. Surprising exceptions persist
semiclassically in certain anti-de Sitter (AdS) black holes
\cite{Dias:2019ery,Balasubramanian:2019qwk,Papadodimas:2019msp}, but these do
not seem to survive small changes to the theory or background
\cite{Emparan:2020rnp,Kolanowski:2023hvh}.

Using our disentanglement criteria, we will now provide an exceedingly general
study of SCC in black holes of arbitrary spacetimes. Previously,
ref.~\cite{Papadodimas:2019msp} derived entanglement criteria necessitated by
smooth Cauchy horizons that can be evaluated using free field modes on a given
background. In contrast, we will argue in a more physically transparent way for
our disentanglement criteria, without reliance on any explicit calculations and
being largely insensitive to details of the background. Cauchy horizons
satisfying our disentanglement criteria not only fail to be smooth, but are
sufficiently singular to enforce SCC, as we now explain.

Let us demonstrate SCC by contradiction: suppose that there is some sensible
candidate extension beyond the Cauchy horizon. Then, we may consider the mutual
information \(I\) between the regions \(R_1\) and \(R_2\) shown respectively in
grey and yellow in \cref{fig:cauchyhorizon}. As illustrated there, we deform
\(R_1\) and \(R_2\) such that the black entangling surface between them is
pushed along the blue null congruence through the red Cauchy horizon. We will
soon argue that, with the possible exception of special cases in AdS, our
disentanglement criteria \labelcref{item:disentanglement3} and thus
\labelcref{item:disentanglement1} are satisfied at the intersection
\(\lambda_0\) of the null congruence with the Cauchy horizon (\ie{}, the middle
black dots in \cref{fig:cauchyhorizon}). As part of being ``sensible'', we
expect that the geometry and state are not persistently singular above the
Cauchy horizon, so that the regularity conditions
\labelcref{item:smoothnessafter1,item:smoothnessafter2} are eventually met.
However, the QNEC and QFC then imply the presence of infinite energy and a
strong spacetime singularity, in violation of conditions
\labelcref{item:finiteenergy,item:finiteexpansion,item:finitearea}. As
previously described, such singularities are too severe to recover (even weak)
solutions to the gravitational equations of motion, thus enforcing
SCC.\footnote{It is perhaps worth emphasizing that our argument is not
  constructive, but rather proceeds by contradiction. Thus, what we learn is not
  necessarily that conditions
  \labelcref{item:finiteenergy,item:finiteexpansion,item:finitearea} fail in the
  \emph{true} evolution of the system. Instead, the lesson is that these
  conditions are inconsistent with the \emph{hypothetical} extension beyond the
  Cauchy horizon satisfying the mild regularity conditions
  \labelcref{item:smoothnessafter1,item:smoothnessafter2}.

  To address the caveat about the QNEC mentioned at the end of
  \cref{sec:qnecqfc}, one might initially hypothesize that perturbation theory
  can be applied by starting from a classical background that is suitably weakly
  curved near the Cauchy horizon. The argument by contradiction would then
  falsify this hypothesis or the other, stated conditions. In this sense,
  possible violations of SCC are excluded from the perturbative semiclassical
  regime. (Again, let us note that the argument following from the QFC can
  proceed independently from the QNEC and is not subject to the aforementioned
  caveat.)}

\section{Disentanglement across the Cauchy horizon}
\label{sec:cauchyhorizons}
It remains to argue that the mutual information \(I_{R_1:R_2}\) does indeed
become \(-\infty\) across the entangling surface \(\lambda_0\) on the Cauchy
horizon. To make any progress, we must confront the uncertain physics introduced
by the high curvature region near the naive black hole singularity, drawn as
zigzags in \cref{fig:cauchyhorizon}. However mysterious the dynamics may be, we
nonetheless expect there to be some evolution --- more precisely, a quantum
channel --- relating the state on the portion \(R_3\) of the Cauchy horizon
below \(\lambda_0\) to the state on \(R_2\). In \cref{fig:cauchyhorizon}, we
have drawn \(R_3\) in green. This precludes, for example, the perverse
possibility for the black hole singularity to magically re-emit into \(R_2\)
bits of quantum information previously absorbed by the future asymptotic
boundary \(\mathscr{I}^+\). Certainly, our knowledge of semiclassical gravity
may be insufficient to \emph{determine} the evolution \(R_3\to R_2\); we are
assuming merely that such a channel \emph{exists}. Since mutual information is
monotonic \(I_{R_1:R_3} \ge I_{R_1:R_2}\) through quantum channels, it will
suffice to argue that \(I_{R_1:R_3}\) is negatively divergent.\footnote{Below,
  we present an argument based on physical intuition about the entanglement
  carried by QFT degrees of freedom. In \cref{sec:cftonbh}, we explicitly
  calculate the mutual information \(I_{R_1:R_3}\) for two-dimensional CFTs in
  arbitrary stationary black holes. The conclusions of our intuitive argument
  below for asymptotically flat, dS, and AdS black holes are all precisely
  reproduced. We also calculate the CFT stress tensor and compare mutual
  information to this more standard probe of SCC.}

The culprits behind this entanglement deficit are the same infinitely many,
arbitrarily energetic degrees of freedom \(r_1\subset R_1\) that Penrose
prophetically identified in his original argument.\footnote{When the fields are
  free in the ultraviolet, these degrees of freedom can be precisely identified
  as the local Rindler-like modes constructed by
  ref.~\cite{Papadodimas:2019msp}, which we describe in \cref{foot:rindler}. The
  important point is not only that there are infinitely many of these modes, but
  also that they would ordinarily, in a smooth state, make an infinite
  contribution to the mutual information between adjacent subregions.

  An interesting point is that this a logarithmic infinity, as mentioned in
  \cref{foot:rindler}, such that the disentanglement of these modes results in a
  deficit in mutual information which diverges logarithmically in the affine
  separation \(\lambda_0-\lambda\) between the Cauchy horizon and the entangling
  surface. Meanwhile, in more standard approaches to probing SCC, the stress
  tensor has been found to diverge quadratically with respect to the Kruskal
  coordinate across the inner horizon, sometimes negatively --- see, \eg{}
  ref.~\cite{Hollands:2019whz}. The strength of this divergence is therefore
  compatible with the QNEC \labelcref{eq:qnec_mi} and the logarithmic
  entanglement deficit of the Rindler-like modes. In
  \cref{sec:energyentanglement}, studying two-dimensional CFTs, we
  quantitatively examine this relationship between mutual information and the
  stress tensor near the Cauchy horizon. \label{foot:specificmodes}} Here,
however, these degrees of freedom need not be excited outside the black hole
from the local vacuum as in the classical case. As we will elaborate, what
concerns us is that the infinitely many degrees of freedom in \(r_1\) do not
seem to be entangled with degrees of freedom in \(R_3\) and thus \(R_2\). This
can be contrasted with analogous degrees of freedom (dashed wavy arrows in
\cref{fig:cauchyhorizon}) entangled across earlier choices of the entangling
surface \(\lambda_{\alpha<0}\), where they make an infinite positive
contribution (balanced by counterterms) in \(I_{R_1:R_2}[\lambda_{\alpha<0}]\).
We will now argue that it is impossible to make up for the potentially missing
analogous entanglement across \(\lambda_0\), except in some AdS cases.

\begin{figure}
  \hfill
  \begin{subfigure}[t]{0.3\textwidth}
    \begin{center}
      \includegraphics[scale=1.25]{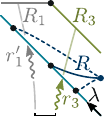}
    \end{center}
  \end{subfigure}
  \hfill
  \begin{subfigure}[t]{0.3\textwidth}
    \begin{center}
      \includegraphics[scale=1.25]{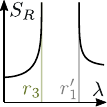}
    \end{center}
  \end{subfigure}
  \hfill\null
  \caption{\label{fig:excitedr} Infinitely entangled excitations \(r_1'\) and
    \(r_3\) produce divergences in entropy \(S_R\).}
\end{figure}

Let us proceed by supposing the contrary, that the entanglement budget is
balanced by infinite entanglement between some degrees of freedom \(r_1'\subset
R_1\) and \(r_3\subset R_3\) so that \(I_{R_1:R_3} > -\infty\). \textit{A
  priori}, \(r_1'\) need not be the same as \(r_1\). However, any such \(r_1'\ne
r_1\) must constitute infinite energy excitations, as we now explain by tracking
these degrees of freedom back in time. The infinite entanglement carried by
\(r_1'\), which compensates for \(r_1\) at \(R_1\), must also contribute to the
renormalized entropies of earlier subregions in the interior of the spacetime
region away from the Cauchy horizon --- we illustrate this in
\cref{fig:excitedr}. As shown, we may apply the QNEC and QFC to the light-blue
null congruence which crosses the path of these excitations (with entropy \(2S\)
replacing mutual information \(I\) in \cref{eq:qnec_mi,eq:qfc_mi}). Due to the
infinite entanglement carried by the excitations, the QNEC and QFC imply that
they must also carry infinite energy and thus regrettably imprint a strong
singularity on spacetime. To keep the spacetime intact up to the Cauchy horizon,
we must therefore make do with \(r_1'=r_1\).

The above considerations highlight a more general point: at least prior to
approaching the Cauchy horizon, the infinite entanglement between \(r_1'=r_1\)
and \(r_3\) must be attributed to the correlations in locally vacuum-like
fluctuations, as opposed to excitations. As previously mentioned, the former
type of entanglement is renormalized by counterterms in entropy and mutual
information, so the disastrous conclusions drawn from the QNEC and QFC in the
previous paragraph would be nullified. However, due to the short-distance nature
of such entanglement, (nearly all of) the degrees of freedom \(r_1\) and \(r_3\)
must have travelled closely together up to the Cauchy horizon. The final,
kinematical puzzle is how this can be achieved, given the ultrarelativstically
infalling momenta carried by \(r_1\), and the parametrically different
momenta\footnote{As a toy analysis, one can consider timelike and null geodesics
  in a static black hole as geometric approximations to the trajectories of
  localized free field modes. Considering the sign of the conserved quantity
  associated to the static Killing vector, it is easy to see that no null or
  timelike geodesics connect the (right) event horizon to the (right) Cauchy
  horizon. In other words, a particle in \cref{fig:cauchyhorizon} that fell
  across the event horizon, \eg{} closely following \(r_1\), will carry too much
  inertia to freely cross \(R_3\). Such a simple argument may not exist more
  generally, but the degrees of freedom \(r_1\) always carry parametrically
  large infalling momenta --- see \cref{foot:rindler,foot:specificmodes} ---
  such that they should not cross the Cauchy horizon arbitrarily shortly above
  \(\lambda_0\).} of \(r_3\) required to cross \(R_3\).

With the absence of any conceivable resolutions in asymptotically flat and dS
spacetimes, the argument is complete in these cases. In fact, the dS argument
can be further strengthened. Tracking \(r_1\) back in time, as illustrated in
\cref{fig:cauchyhorizon_ds}, we realize that \(r_1\) must be entangled across
the cosmological horizon to degrees of freedom captured by \(\mathscr{I}^+\) ---
otherwise, we would be running a disentanglement argument for a singular
cosmological horizon instead of the Cauchy horizon. Calculations with free
fields also show this explicitly for smooth states \cite{Shrivastava:2020xmw}.
What remains unappreciated\footnote{I thank Stefan Hollands for discussing this
  point with me.}, however, is the monogamous nature of entanglement, which
forbids \(r_1\) from further entangling with other partners \(r_3\). Thus, the
putative entanglement between \(r_1\) and \(r_3\) is forbidden by both
kinematics and entanglement monogamy.\footnote{For two-dimensional CFTs, we show
  in \cref{sec:energyentanglement} that \(I_{R_1:R_3}\) indeed diverges
  negatively in asymptotically flat and dS black holes. While the coefficient of
  the divergence in the stress tensor vanishes when the temperature of the
  infalling (cosmological)radiation matches the inner horizon temperature
  \cite{Hollands:2019whz}, the divergence in \(I_{R_1:R_3}\) is unaffected by
  the infalling radiation. This supports the above discussion about the capture
  of entangled partners by \(\mathscr{I}^+\)and illustrates how mutual
  information can serve as a more refined probe of SCC than local observables .}

In contrast, the AdS asymptotic boundary \(\mathscr{I}\) allows the degrees of
freedom \(r_1\) to dramatically change their kinematics by bouncing
off.\footnote{This paragraph is about reflecting boundary conditions at the AdS
  boundary \(\mathscr{I}\). Instead, one may, for example, couple the
  asymptotically AdS spacetime to a bath, where \(\mathscr{I}\) is a transparent
  interface between the two systems
  \cite{Almheiri:2019psf,Almheiri:2019yqk,Chen:2019uhq,Chen:2020jvn}. In such
  cases where outgoing degrees of freedom are not reflected into the black hole,
  the situation is more akin to an asymptotically flat or dS black hole, so we
  do expect the Cauchy horizon to satisfy our disentanglement criteria.}
Tracking \(r_1\) back in time, we see in \cref{fig:cauchyhorizon_ads} that these
degrees of freedom were formerly Hawking radiation emitted near the black hole
event horizon. Again, to avoid an earlier singularity here, they must be
entangled with partners across the horizon --- this is famously connected to the
information paradox \cite{Almheiri:2012rt}. But in contrast to dS, we see in
\cref{fig:cauchyhorizon_ads} that these partners land in \(R_3\) and we may be
able to identify them as the sought after \(r_3\). Thus, our analysis remains
inconclusive for AdS, which is a somewhat comforting check that we have not
cheated.\footnote{In \cref{sec:energyentanglement}, we calculate \(I_{R_1:R_3}\)
  for two-dimensional CFTs in asymptotically AdS black holes. The mutual
  information can diverge with either sign depending on the inner and outer
  horizon temperatures. While our disentanglement criteria only hold for a range
  of temperatures, positively divergent mutual information and entropies are
  also dangerous due to the QNEC and QFC. It would be worthwhile to obtain a
  general explanation of when disentanglement or hyper-entanglement occurs at
  the Cauchy horizons in AdS black holes, with the hope of generically securing
  SCC. (However, as ref.~\cite{Papadodimas:2019msp} points out, the fate of SCC
  in AdS black holes may yet be obscured by late-time phenomena such as
  tunnelling \via{} wormhole emission \cite{Stanford:2022fdt}.)} After all,
strong singularities appear to be absent at the Cauchy horizons in some special
AdS black holes \cite{Dias:2019ery, Balasubramanian:2019qwk,
  Papadodimas:2019msp} (but apparently reappear with slight changes to the
theory or background \cite{Emparan:2020rnp,Kolanowski:2023hvh}).

\section{Conclusion}
By appealing to the QNEC and QFC, we have concretely related the destruction of
entanglement to the presence of infinite energies and strong spacetime
singularities. The power of these ideas was demonstrated by giving an
exceedingly general and physically transparent explanation of SCC for quantum
fields in semiclassical black holes. While the causal structure of spacetime
clearly dictates the flow of entanglement, we have conversely seen how regions
starved of entanglement are rejected from spacetime. It is intriguing that
gravity is so inextricably codependent on quantum information.

\section*{Acknowledgements}
I am grateful to Stefan Hollands, Gary T. Horowitz, Maciej Kolanowski, Donald
Marolf, Arvin Shahbazi-Moghaddam, and Marija Toma\v{s}evi\'c for helpful
discussions. I would like to thank the Institute for Advanced Study for its
support and hospitality during the Workshop on Spacetime and Quantum Information
held on December 11--13, 2023. I am supported by a Fundamental Physics Fellowship
through the University of California, Santa Barbara.

\appendix

\section{Energy and entanglement in two-dimensional CFTs}

In the intuitive spirit of this essay, thus far, we have maintained a rather
qualitative level of discussion about the entanglement and disentanglement
between QFT degrees of freedom. The goal of this appendix is to supplement the
intuition we have developed in the main body of this essay with quantitative
examples. For simplicity, we reduce attention to conformal field theories (CFTs)
in two spacetime dimensions.

While the metric has no propagating degrees of freedom in two dimensions, one
can consider toy models of gravity, \eg{} Jackiw-Teitelboim (JT) gravity, by
introducing other ``geometric'' degrees of freedom, \eg{} a dilaton. The role
played by area in connection to entropy and the QFC is then taken up by these
degrees of freedom, so our analysis in \cref{sec:disentanglement} should have
generalizations to such toy models\footnote{JT gravity has been extensively
  dissected as a toy model for gravity in the context of the black hole
  information paradox, as initiated by \cite{Almheiri:2019psf}. The QFC has been
  used in this context to avoid some auxiliary paradoxes
  \cite{Almheiri:2019yqk}. See \cite{Kolanowski:2023hvh} for a discussion of
  two-dimensional dilaton theories with respect to SCC.}. Since our focus is on
field theory in this appendix, however, we will consider only CFTs on fixed
background metrics.

In these two-dimensional CFTs, we will be able to calculate the stress tensor,
entropy, and mutual information explicitly. Leveraging conformal and Weyl
transformations, we will be able to perform these calculations while being
completely agnostic to the chosen CFT --- it may be free or strongly
interacting, contain one or many fields. In \cref{sec:joiningquench}, we review
the example of a joining quench, where the infinite energy accompanying
disentanglement can be calculated in a regulated manner. Next, in
\cref{sec:cftonbh}, we examine CFTs on arbitrary stationary black hole
backgrounds. We evaluate the mutual information \(I_{R_1:R_3}\) for the regions
\(R_1\) and \(R_3\) which played starring roles in our argument for
disentanglement across Cauchy horizons in \cref{sec:cauchyhorizons}. This will
provide opportunities to compare our approach with other approaches to probing
SCC based on divergences in the stress tensor.

\subsection{A joining quench example}
\label{sec:joiningquench}

\begin{figure}
  \begin{subfigure}{0.45\textwidth}
    \begin{center}
      \includegraphics[scale=1.25]{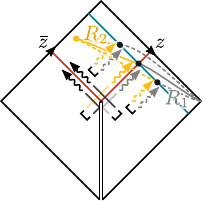}
    \end{center}
    \caption{\label{fig:joiningquench}Two half-spaces, initially in their vacuum
      states and separated by a wall, are suddenly joined. The joint produces
      null energy shock waves (red), across which certain degrees of freedom
      (solid grey and yellow wavy arrows) are entangled not with each other, but
      far away partners (black wavy arrows).}
  \end{subfigure}
  \hfill
  \begin{subfigure}{0.45\textwidth}
    \begin{center}
      \includegraphics[scale=1.25]{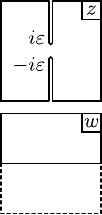}
    \end{center}
    \caption{\label{fig:joiningquenchreg}The CFT state can be prepared by a
      Euclidean path integral over the geometry in the top panel, which can then
      be conformally mapped by \cref{eq:conftrans} to the upper-half-plane in
      the lower panel.}
  \end{subfigure}
  \par\bigskip
  \begin{subfigure}[t]{0.45\textwidth}
    \begin{center}
      \includegraphics[scale=1.25]{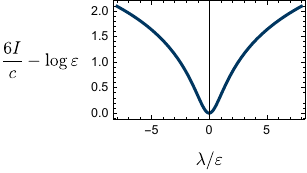}
    \end{center}
    \caption{\label{fig:joiningquench_mi}Renormalized mutual information
      \(I_{R_1:R_2}\) plotted against the affine parameter \(\lambda=-z_\ent\)
      at the entangling surface \((z_\ent,\bar{z}_\ent)\) along the blue null
      line in (\subref{fig:joiningquench}). The yellow endpoint in
      (\subref{fig:joiningquench}) is chosen to be at \(\bar{z}_2=1/2\), while
      the plot is independent of the choices of \(\bar{z}_\ent\) and \(z_2\).}
  \end{subfigure}
  \hfill
  \begin{subfigure}[t]{0.45\textwidth}
    \begin{center}
      \includegraphics[scale=1.25]{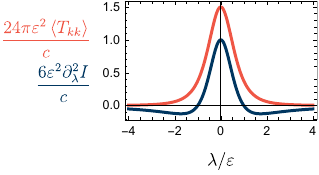}
    \end{center}
    \caption{\label{fig:joiningquench_mi2ndDeriv}The stress tensor expectation
      value and the second derivative of mutual information. The QNEC
      \labelcref{eq:qnec_2D} states that the red curve is bounded from below by
      the dark blue curve.}
  \end{subfigure}
  \caption{Disentanglement from a joining quench.}
\end{figure}

As shown in \cref{fig:joiningquench}, we consider a joining quench of two
half-spaces \cite{Calabrese:2007mtj, Ugajin:2013xxa,
  Shimaji:2018czt}\footnote{Joining quenches have also played a part in
  considerations of the black hole information paradox, where an AdS black hole
  is joined to a bath which collects Hawking radiation. See, \eg{}
  \cite{Almheiri:2019psf}, which also provides a reasonable review of CFT
  entanglement entropy calculations as well as the connection to twist
  operators, boundaries, conformal transformations, and Weyl transformations ---
  ideas which we will make use of.}. In the lower half of
\cref{fig:joiningquench}, the two half-spaces --- \(\mathbb{R}_{<0}\) and
\(\mathbb{R}_{>0}\) respectively \(\times\)time --- are initially unentangled
and separated by a wall where reflecting boundary conditions are imposed. For
concreteness, let us prepare each of the half-spaces in the vacuum state, where
the stress tensor \(\ev{T_{\mu\nu}} = 0\) vanishes. At a given time \(t=0\), the
two half-spaces are then joined suddenly by removing the wall separating them.
We will calculate the stress tensor and mutual information in this more
complicated setup by analytic continuation from Euclidean signature.

In the top panel of \cref{fig:joiningquenchreg}, we have illustrated the flat
Euclidean geometry, with complex coordinates \((z,\bar{z})\), over which a path
integral prepares, up to a regulator \(\varepsilon\), the aforementioned
half-space vacua product state on the \((-z+\bar{z})/2=0\) time slice. As shown,
an extra \(\varepsilon\) amount of Euclidean evolution has been applied to the
state, with the wall removed. As we will see, this regulates the otherwise
infinite energy of the joined state by exponentially \(e^{-\varepsilon E}\)
suppressing high energy \(E\) eigenstates of the joined, \ie{} Minkowski,
Hamiltonian. Moreover, it has the effect of reconstructing the Minkowski vacuum
entanglement between degrees of freedom in the two half-spaces at energies
\(\gtrsim 1/\varepsilon\), while leaving lower energy degrees of freedom
untouched in the half-space vacua.

The conformal transformation
\begin{align}
  w &= i\sqrt{\frac{i\varepsilon+z}{i\varepsilon-z}}
      \label{eq:conftrans}
\end{align}
maps the geometry in the top panel of \cref{fig:joiningquenchreg} to the \(w\)
upper-half-plane, as shown in the bottom panel. In this half-space geometry, the
stress tensor again vanishes. Moreover, as we will describe further below,
entanglement entropies and mutual information can be evaluated using the
half-space correlation functions of twist operators. These answers can then be
conformally mapped through \cref{eq:conftrans} and analytically continued to the
Lorentzian spacetime in \cref{fig:joiningquench}, where \((z,\bar{z})\) are null
coordinates.

The resulting stress tensor is\footnote{A direct analytic continuation, as
  written here, describes a time-reversal-symmetric Lorentzian evolution where
  the wall is removed to both the future and the past. This differs from the
  physical evolution illustrated in \cref{fig:joiningquench} only in the causal
  past of the joint \(z=\bar{z}=0\) (in the \(\varepsilon\to 0\) limit).}
\begin{align}
  \ev{T_{zz}}
  &= - \frac{c}{24\pi} \{w,z\}
    = \frac{c\, \varepsilon^2}{16\pi(z^2+\varepsilon^2)^2}
    \;.
    \label{eq:joiningquench_stress}
\end{align}
Here, \(c\) is the central charge of the CFT, and \(\{w,z\}\) is the Schwarzian
derivative. In the limit \(\varepsilon\to 0\), we see that \(\ev{T_{zz}}\)
becomes a non-integrable contact term \(+\infty\,\delta(z)\) at the right red
horizon in \cref{fig:joiningquench}. The anti-holomorphic component of the
stress tensor similarly reduces to an infinite energy shock wave
\(+\infty\,\delta(\bar{z})\) localized on the left horizon. In addition to
analogous contact terms, the stress tensor for Rindler vacua also has a power
law divergence as one approaches the horizon \cite{Parentani:1993yz}. As we
elaborate below, what our joining quench example illustrates is that
disentanglement need not be accompanied by a stress tensor that diverges
\emph{on approach} to the disentanglement surface as for Rindler vacua; the
infinite energy deduced in \cref{sec:disentanglement} can be localized at the
disentanglement surface.

To quantify disentanglement and study the QNEC, we next consider mutual
information \labelcref{eq:mi} between the grey \(R_1\) and yellow \(R_2\)
regions in \cref{fig:joiningquench} whose black entangling surface (or rather,
point) slides along the blue null line. Let us denote the coordinates of the
yellow left endpoint of \(R_2\) and the black entangling surface respectively by
\((z_2,\bar{z}_2)\) and \((z_\ent,\bar{z}_\ent)\).

As part of the mutual information \(I_{R_1:R_2}\), we will need the entropies
\begin{align}
  S_{R_1}
  &= S^{\uhp}(w(z_\ent),\bar{w}(\bar{z}_\ent))
    - \frac{c}{6} \log \sqrt{w'(z_\ent)\bar{w}'(\bar{z}_\ent)}
    \;,
    \label{eq:joiningquench_entropyr1}
  \\
  S_{R_1\cup R_2}
  &= S^{\uhp}(w(z_2),\bar{w}(\bar{z}_2))
    - \frac{c}{6} \log \sqrt{w'(z_2)\bar{w}'(\bar{z}_2)}
    \;,
    \label{eq:joiningquench_entropyr1r2}
\end{align}
where\footnote{We are considering renormalized entropies here. There bare
  quantities corresponding to \cref{eq:entropy_onepoint,eq:entropy_twopoint}
  have divergences \(\frac{c}{6} \log(\epsilon)\) and
  \(\frac{c}{6}\log(\epsilon^2)\) logarithmic in the ultraviolet cutoff
  \(\epsilon\) of the CFT (not to be confused with the \(\varepsilon\) regulator
  of the shock wave).}
\begin{align}
  S^{\uhp}(w,\bar{w})
  &=
    \frac{c}{6} \log\left(
    \frac{w - \bar{w}}{i}
    \right)
    + \log g
    \label{eq:entropy_onepoint}
\end{align}
derives from the one-point function of a twist operator in the \(w\) half-plane.
The other terms in \cref{eq:joiningquench_entropyr1} result from the conformal
transformation \(w\mapsto z\) mapping this twist insertion to the single
endpoints of the regions \(R_1\) and \(R_1\cup R_2\). In the \(w\) half-plane,
conformal invariance determines one-point functions up to normalization,
manifesting as the boundary entropy \(\log g\) above \cite{Affleck:1991tk} ---
this constant will cancel out below.

Unfortunately, also appearing in \(I_{R_1:R_2}\) is the entropy \(S_{R_2}\) of
the region \(R_2\) with two endpoints, corresponding to a two-point function of
twist operators. While this is not determined by conformal invariance in
general, the limit \(\varepsilon, \abs{z_\ent}\ll \bar{z}_\ent,-z_2,\bar{z}_2\)
where the entangling surface is close to the red horizon in
\cref{fig:joiningquench} corresponds to an operator product expansion (OPE)
limit of the twist operators in the \(w\) half-plane. In this limit, the \(w\)
half-plane two-point function is well approximated by a full-plane two-point
function\footnote{By the method of images, the half-plane two-point function of
  twist operators is similar to a full-plane four-point function. Thus, it is
  controlled by a cross-ratio
  \begin{align}
    \eta(w_1,\bar{w}_1;w_2,\bar{w}_2)
    &= -\frac{(w_1-\bar{w}_1)(w_2-\bar{w}_2)}{(w_1 - w_2)(\bar{w}_1 - \bar{w}_2)}
      \;.
      \label{eq:cratio}
  \end{align}
  (In the present case, the relevant cross-ratio is
  \(\eta(w(z_\ent),\bar{w}(\bar{z}_\ent);w(z_2),\bar{w}(\bar{z}_2))\) where
  \(w\) is given by \cref{eq:conftrans}.) The limit \(\eta \to \infty\)
  corresponds to the OPE limit where the half-plane two-point function of twist
  operators is approximated by a full-plane two-point function and the
  corresponding expression for entropy is \cref{eq:entropy_twopoint}. See, \eg{}
  \cite{Almheiri:2019psf} (but note that we are using a different cross-ratio
  and our boundary is at \(w=\bar{w}\)).
  \label{foot:cratio_ope}}, so that
\begin{align}
  S_{R_2}
  &\sim S^{\pln}(
    w(z_\ent),\bar{w}(\bar{z}_\ent);
    w(z_2),\bar{w}(\bar{z}_2)
    )
    - \frac{c}{6} \log \sqrt{
    w'(z_\ent)\bar{w}'(\bar{z}_\ent)
    w'(z_2)\bar{w}'(\bar{z}_2)
    }
\end{align}
where
\begin{align}
  S^{\pln}(
  w_1,\bar{w}_1;
  w_2,\bar{w}_2
  )
  &= \frac{c}{6} \log \left[
    (w_1 - w_2)
    (\bar{w}_1 - \bar{w}_2)
    \right]
    \;.
    \label{eq:entropy_twopoint}
\end{align}

Collecting the pieces and simplifying using the limit
\(\varepsilon,\abs{z_\ent}\ll \bar{z}_\ent,-z_2,\bar{z}_2\), we find
\begin{align}
  I_{R_1:R_2}
  &\sim \frac{c}{6}\log\left[
    \frac{\sqrt{\varepsilon^2 + z_\ent^2}\, (\bar{z}_\ent - \bar{z}_2) \bar{z}_\ent}{\bar{z}_2}
    \right]
    \;.
    \label{eq:joiningquench_mi}
\end{align}
Using \(k=-\partial_z\) and \(\lambda=-z_\ent\) as the affine generator and
parameter along the blue null line, the two-dimensional QNEC states that
\begin{align}
  \ev{T_{zz}(z_\ent)}
  = \ev{T_{kk}(z_\ent)}
  \ge \frac{1}{4\pi} \partial_\lambda^2 I_{R_1:R_2}
  &= \frac{1}{4\pi} \partial_{z_\ent}^2 I_{R_1:R_2}
    \label{eq:qnec_2D}
  \\
  &\sim \frac{c}{24\pi} \frac{\varepsilon^2 - z_\ent^2}{(\varepsilon^2 + z_\ent^2)^2}
    \;.
    \label{eq:joiningquench_mi2ndDeriv}
\end{align}
We plot
\cref{eq:joiningquench_stress,eq:joiningquench_mi,eq:joiningquench_mi2ndDeriv}
in \cref{fig:joiningquench_mi,fig:joiningquench_mi2ndDeriv}, where we see that
the QNEC is indeed satisfied.

Upon removing the regulator \(\varepsilon\to 0\), the mutual information
\labelcref{eq:joiningquench_mi} is negatively, logarithmically divergent as the
black entangling surface \(\lambda=-z_\ent\) is pushed towards the red horizon
\(\lambda_0=0\), thus satisfying our disentanglement criterion
\labelcref{item:disentanglement1} (and also the stronger criterion
\labelcref{item:disentanglement3}, but gravity is not dynamical here). As
predicted in \cref{sec:disentanglement}, we find infinite energy \(\langle
T_{kk} \rangle = +\infty\,\delta(z)\). In the \(\varepsilon\to 0\) limit,
\cref{eq:joiningquench_mi} is concave-down with respect to \(z_\ent\ne 0\), so
one might naively think that the QNEC \labelcref{eq:qnec_2D} places no bound on
positive energy. However, as made clear in \cref{fig:joiningquench_mi2ndDeriv}
with the \(\varepsilon\) regulator reinstated, \(\partial_\lambda^2
I_{R_1:R_2}\) is large and positive in the region \(-\varepsilon\lesssim z_\ent
\lesssim \varepsilon\), where the QNEC then dictates the presence of positive
energy.

In the spirit of this essay, let us end this example by sketching an intuitive
explanation of the disentanglement across the red horizon. As illustrated in
\cref{fig:joiningquench} for the right horizon, the culprits are right-moving
degrees of freedom (wavy arrows) which are ordinarily (in the dashed cases)
infinitely entangled (as indicated by brackets) at short distances across the
entangling surface (black dot). However, let us consider the analogous degrees
of freedom (solid grey and yellow wavy arrows) across the entangling surface
(middle black dot) located on the horizon. Tracing these back in time, we see
that the sudden removal of the reflecting wall means that the solid grey and
yellow degrees of freedom originate respectively from the right and left
half-spaces and should not be infinitely entangled with each other. Rather,
given the short-distance entanglement present in each half-space vacuum, they
are instead entangled with degrees of freedom (black wavy arrows) which have
propagated away and do not contribute to the entanglement between \(R_1\) and
\(R_2\) as measured by \(I_{R_1:R_2}\).

\subsection{Stationary black hole backgrounds}
\label{sec:cftonbh}

Finally, we consider CFTs on stationary two-dimensional black hole backgrounds,
\begin{align}
  \dd{s}^2
  &= -f(r) \dd{t}^2
    + \frac{\dd{r}^2}{f(r)}
    \;.
    \label{eq:statmetric}
\end{align}
Our goal will be to evaluate the stress tensor and the mutual information for
regions relevant to our disentanglement argument for Cauchy horizons, presented
in \cref{sec:cauchyhorizons}. We begin in \cref{sec:coordinates} by describing
the variety of black holes \labelcref{eq:statmetric} that we will consider and
introduce Kruskal coordinates adapted to various horizons in these geometries.
Then, in \cref{sec:states} we describe the construction of conformal vacua,
corresponding to Unruh and Hartle-Hawking states with no infalling radiation and
infalling radiation at various temperatures. We also describe the compatibility
of these vacua with the asymptotically flat, dS, and AdS geometries we consider.
Finally, in \cref{sec:energyentanglement}, we evaluate the stress tensor and
mutual information in these geometries and states.

\subsubsection{The geometry and useful coordinates}
\label{sec:coordinates}

The curvature of the metric \labelcref{eq:statmetric} reads
\begin{align}
  R &= -f''(r)
      \;,
      \label{eq:statricci}
\end{align}
so the \(\sim r^2\) asymptotics of \(f\) at large \(r\) determine whether the
spacetime is asymptotically flat, dS, or AdS at its boundary \(r=+\infty\). For
relevance to SCC, we consider black holes possessing Cauchy, inner horizons
\(r=r_->0\) in addition to event, outer horizons \(r=r_+>r_-\). For
asymptotically dS geometries, there will also be cosmological horizons
\(r=r_\cosm>r_+\). The blackening factor \(f\) then satisfies
\begin{align}
  f(r<r_-)
  &> 0 \;,
  &
    f(r_-<r<r_+)
  &< 0 \;,
  &
    f(r_+<r<r_c)
  &> 0 \;,
  &
    f(r_\cosm<r)
  &< 0
\end{align}
and vanishes on the horizons,
\begin{align}
  f
  &= \begin{cases}
    2 \kappa_+ (r - r_+)
    + \order{(r - r_+)^2}
    \\
    2 \kappa_- (r_- - r)
    + \order{(r - r_-)^2}
    \\
    2 \kappa_\cosm (r_\cosm - r)
    + \order{(r-r_\cosm)^2}
  \end{cases}
    \label{eq:blackening_nearhorizon}
\end{align}
where the \(\kappa_{\pm/\cosm}\) denote the respective horizon temperatures. We
will not need to specify \(f\) any further.\footnote{There is one more
  assumption that is needed for the size of the corrections in the stress
  tensors written below in
  \cref{eq:flatunruh_stressenergy,eq:dsunruh_stressenergy,eq:hh_stressenergy}.
  We assume that \(f\) admits an expansion near the inner horizon,
  \begin{align}
    f &=2 a_- \kappa_- z_- \bar{z}_- + b_-\, (z_- \bar{z}_-)^2 + \order{(z_- \bar{z}_-)^3}
        \label{eq:blackening_nearhorizon_extra}
        \;,
  \end{align}
  in terms of the Kruskal coordinates \((z_-,\bar{z}_-)\) and constant \(a_-\)
  introduced below, and another constant \(b_-\). The leading order piece is
  already expressed in \cref{eq:blackening_nearhorizon}. The additional
  assumption is that the next-to-leading order correction is precisely quadratic
  and this approximation is accurate up to \(\order{(z_- \bar{z}_-)^3}\)
  corrections. \label{foot:blackening_nearhorizon_extra}}

Let us clarify that we are not specializing only to dS. Relations involving
\(r_\cosm\) are only relevant for dS and should otherwise be ignored or modified
in the obvious way. In cases other than dS, \(\kappa_\cosm\) will be thought of
below as a free parameter, not necessarily determined by geometry as in
\cref{eq:blackening_nearhorizon}. We will later see that \(\kappa_\cosm/2\pi\)
describes the temperature of ``cosmological'' infalling radiation in certain
states, whether that be on dS or asymptotically flat spacetimes.

Let us introduce some useful coordinates next. The radial tortoise coordinate
\(r_*\) is defined by
\begin{align}
  \dd{r_*}
  &= \frac{\dd{r}}{f}
    \;,
\end{align}
up to an integration constant that we will not specify --- it merely determines
the \(a_\pm\), \(a_\cosm\), and \(R_{\mathscr{I}}\) constants below, which we
choose to be all positive. The near-horizon behaviour of \(r_*\)
reads\footnote{Due to the logarithms, \(\Im r_*\) jumps discontinuously across
  each horizon; there are also jumps in \(\Im t\) such that the Kruskal
  coordinates below are defined consistently with respect to \(t\) and \(r_*\).
  We will not dwell on these details, but see \eg{}
  \cite{Balasubramanian:2019qwk} for discussion of these imaginary shifts in
  relation to the Kubo–Martin–Schwinger condition on thermal states. Closely
  related are \cref{eq:outerkms_z,eq:outerkms_t} below.}
\begin{align}
  r_*
  &= \begin{cases}
    \frac{1}{2\kappa_+}
    \log\left( \frac{r-r_+}{a_+} \right)
    + \order{r-r_+}
    \\
    -\frac{1}{2\kappa_-} \log\left( \frac{r - r_-}{a_-} \right)
    + \frac{i \pi}{2\kappa_+}
    + \order{r-r_-}
    \\
    -\frac{1}{2\kappa_\cosm} \log\left( \frac{r_c - r}{a_\cosm} \right)
    + \order{r-r_\cosm}
  \end{cases}
    \;.
\end{align}

\begin{figure}
  \begin{subfigure}{0.45\textwidth}
    \begin{center}
      \includegraphics[scale=1.25]{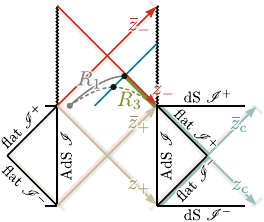}
    \end{center}
    \caption{\label{fig:stationarylor}Lorentzian black holes. The real ranges
      \((-\infty,+\infty)\) of the various \(z\) and \(\bar{z}\) Kruskal
      coordinates are indicated by lines tipped by arrows at \(+\infty\). The
      tortoise null coordinates \(z_\tor\) and \(\bar{z}_\tor\) are not
      illustrated; their real ranges correspond respectively to the positive
      ranges of \(z_+\) and \(\bar{z}_+\).}
  \end{subfigure}
  \hfill
  \begin{subfigure}{0.45\textwidth}
    \begin{center}
      \includegraphics[scale=1.25]{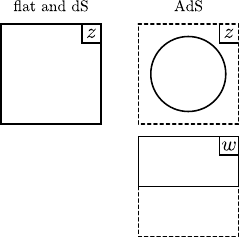}
    \end{center}
    \caption{\label{fig:stationaryeuc}The asymptotically flat and dS vacua we
      consider are prepared on a Euclidean plane. The AdS Hartle-Hawking vacuum
      is prepared on a disk which can be conformally mapped to the half-plane.}
  \end{subfigure}
  \caption{Two-dimensional stationary black holes and the preparation of CFT
    states.}
\end{figure}

From \(t\) and \(r_*\) we can define tortoise null coordinates
\begin{align}
  z_\tor
  &= r_* - t
    \;,
  &
    \bar{z}_\tor
  &= r_* + t
    \;,
    \label{eq:tornull}
\end{align}
and Kruskal coordinates\footnote{Our null tortoise and Kruskal coordinates are
  related to the more standard ones, often called \(u\), \(v\), \(U\), and
  \(V\), by \(z_\tor=-u\), \(\bar{z}_\tor=v\), \(z_{\pm/\cosm}=-U_{\pm/\cosm}\),
  and \(\bar{z}_{\pm/\cosm}=V_{\pm/\cosm}\). The reasoning behind our
  conventions will become apparent when we discuss Euclidean state preparation
  in \cref{sec:states}.}, \eg{} adapted to the outer horizon,
\begin{align}
  z_+
  &= e^{\kappa_+ z_\tor}
    \;,
  &
    \bar{z}_+
  &= e^{\kappa_+ \bar{z}_\tor}
    \;.
    \label{eq:kruskalouter}
\end{align}
There are similar Kruskal coordinates adapted to the inner and cosmological
horizons (though not given directly by a substitution of subscripts in
\cref{eq:kruskalouter}). The ranges of various null coordinates are illustrated
in \cref{fig:stationarylor}. Where the real ranges of any two \(z\) or any two
\(\bar{z}\) Kruskal coordinates overlap in spacetime, the coordinates are
related by
\begin{align}
  (-z_+)^{1/\kappa_+}
  &= z_-^{-1/\kappa_-}
    \;,
  &
    \bar{z}_+^{1/\kappa_+}
  &= (-\bar{z}_-)^{-1/\kappa_-}
    \;,
  \\
  z_+^{1/\kappa_+}
  &= (-z_\cosm)^{-1/\kappa_\cosm}
    \;,
  &
    \bar{z}_+^{1/\kappa_+}
  &= (-\bar{z}_\cosm)^{-1/\kappa_\cosm}
    \;,
  \\
  & &
      (-\bar{z}_-)^{1/\kappa_-}
  &= (-\bar{z}_\cosm)^{1/\kappa_\cosm}
    \;.
\end{align}
In the case of AdS, the asymptotic boundary is located at finite \(r_*\) and
thus finite
\begin{align}
  \eval{z_+\bar{z}_+}_{\text{AdS \(\mathscr{I}\)}}
  &= R_{\mathscr{I}}^2
    \;.
    \label{eq:adsboundary}
\end{align}

The black hole metric \labelcref{eq:statmetric} becomes conformally flat when
written in terms of null coordinates,
\begin{align}
  \dd{s}^2
  &= \frac{\dd{z_I} \dd{\bar{z}_{J}}}{\Omega_{IJ}^2} \;.
  &
  &(I,J\in\{\tor,+,-,\cosm\}\text{ not summed})
    \label{eq:statmetric_weyl}
\end{align}
Of particular interest to us will be the following coordinate pairings,
\begin{align}
  \Omega_{+\,\tor}^{-2}
  &= \frac{f}{\kappa_+ z_+}
    = - \frac{2 a_- \kappa_-}{\kappa_+}
    z_-^{1+ \frac{\kappa_+}{\kappa_-}}
    \bar{z}_-
    \left[1+\order{z_- \bar{z}_-}\right]
    \;,
    \label{eq:weyl_flatunruh}
  \\
  \Omega_{+\,\cosm}^{-2}
  &= -\frac{f}{\kappa_+ \kappa_\cosm z_+ \bar{z}_\cosm}
    = \frac{2a_- \kappa_-}{\kappa_+ \kappa_\cosm}
    z_-^{1+\frac{\kappa_+}{\kappa_-}}
    (-\bar{z}_-)^{1-\frac{\kappa_\cosm}{\kappa_-}}
    \left[1+\order{z_- \bar{z}_-}\right]
    \;,
    \label{eq:weyl_dsunruh}
  \\
  \Omega_{+ +}^{-2}
  &= \frac{f}{\kappa_+^2 z_+ \bar{z}_+}
    = \frac{2 a_- \kappa_-}{\kappa_+^2}
    z_-^{1+\frac{\kappa_+}{\kappa_-}}
    (-\bar{z}_-)^{1+\frac{\kappa_+}{\kappa_-}}
    \left[1+\order{z_- \bar{z}_-}\right]
    \;.
    \label{eq:weyl_hh}
\end{align}
Above, we have indicated the behaviours of these Weyl factors as the inner
horizon is approached from below, expressed in terms of the inner horizon
Kruskal coordinates.

\subsubsection{State preparation}
\label{sec:states}

We turn now to the question of what state we should consider in the CFT. For
simplicity, we shall consider three choices of conformal vacua, specified below.
However, we also expect that more general states excited above these vacua will
behave qualitatively similarly. For example, in a free theory, the authors of
\cite{Hollands:2019whz} considered the difference in the stress tensor
expectation value between an Unruh state and an arbitrary other state that is
also Hadamard in the same spacetime region. It was shown that this difference
has the same regularity as a classical stress tensor when approaching the Cauchy
horizon in a black hole. Thus, for the purposes of studying quantum enhancements
to SCC, the authors focused on the Unruh state. In similar spirit, we expect the
stress tensor and entanglement structure for the CFT states constructed below to
also be representative of more general states, at least to leading order near
the Cauchy horizon.

Let us now describe the general strategy for constructing our conformal vacua.
First, we remove the Weyl factor in the metric \cref{eq:statmetric_weyl} so that
it becomes flat \(\dd{z}\dd{\bar{z}}\). Next, we continue this flattened
geometry to Euclidean signature, where \(z\) and \(\bar{z}\) now play the role
of holomorphic and anti-holomorphic coordinates over a complex plane or a subset
of it, as illustrated in \cref{fig:stationaryeuc} and elaborated below. The
vacuum state can then be prepared by a path integral over this Euclidean
geometry, where we can calculate quantities such as the stress tensor
expectation value and entropies. Finally, we continue back to Lorentzian
signature and reinstate the Weyl factor in the metric while applying the
corresponding Weyl transformations to the quantities we have calculated. The key
choice in this procedure is the pair of coordinates \(z\) and \(\bar{z}\) ---
different selections from the various null coordinates we have previously
identified will result in different vacua.

\paragraph{Hartle-Hawking state at the outer horizon temperature.} One choice is
the outer horizon Kruskal coordinates \((z_+,\bar{z}_+)\), which results in the
Hartle-Hawking state of the black hole. From \cref{eq:tornull,eq:kruskalouter},
we see that the statement
\begin{align}
  (z_+,\bar{z}_+) \sim (e^{2\pi i} z_+, e^{-2\pi i}\bar{z}_+)
  \label{eq:outerkms_z}
\end{align}
in the Euclidean complex plane corresponds to the identification of time in the
thermal circle at the temperature \(\kappa_+/2\pi\) of the outer horizon
\begin{align}
  i t
  &\sim i t + \frac{2\pi}{\kappa_+}
    \;.
    \label{eq:outerkms_t}
\end{align}
This Hartle-Hawking state is therefore a thermofield double of the left and
right exteriors at the outer horizon temperature. The geometry of the complex
plane is smooth and, in particular, has no conical singularity at the
bifurcation point \(z_+=\bar{z}_+=0\), due to \cref{eq:outerkms_z}. Continuing
to Lorentzian signature, the Hartle-Hawking vacuum remains smooth over the null
ranges covered by the \((z_+,\bar{z}_+)\) Kruskal coordinates. In particular, it
is smooth on the maximally extended outer horizon.

Moreover, for asymptotically flat and AdS black holes, this vacuum is smooth
over any Cauchy surface (away from the inner horizon) running between the
spatial infinities of the left and right exteriors. Among the three vacua we
consider, the Hartle-Hawking vacuum is the only one consistent with the
reflecting boundary conditions at the asymptotic boundary \(\mathscr{I}\) of
AdS. The right- and left-moving, \ie{} holomorphic and anti-holomorphic, degrees
of freedom are related here and not independent. Recalling that \(\mathscr{I}\)
is described by \cref{eq:adsboundary}, the continuation of the AdS black hole to
Euclidean signature is a disk in the complex plane with radius
\(R_{\mathscr{I}}\). As illustrated on the right in \cref{fig:stationaryeuc} and
similar to what we did in \cref{sec:joiningquench}, it will be helpful for
entropy calculations to consider an additional, Mobius transformation to the
upper half plane,
\begin{align}
  w
  &= -i\, \frac{z_+ + i R_{\mathscr{I}}}{z_+ - i R_{\mathscr{I}}}
    \;,
  &
    \bar{w}
  &= i\, \frac{\bar{z}_+ - i R_{\mathscr{I}}}{\bar{z}_+ + i R_{\mathscr{I}}}
    \;.
    \label{eq:disktoplane}
\end{align}
For dS, except when the outer horizon \(\kappa_+/2\pi\) and cosmological
\(\kappa_-/2\pi\) temperatures match, the Hartle-Hawking state, constructed from
\((z_+,\bar{z}_+)\), will be singular on the cosmological horizon. When the
temperatures do match, the Hartle-Hawking state is the same as the Unruh state
we consider below.

\paragraph{Unruh state with infalling cosmological radiation.}
Applying a similar construction to the cosmological Kruskal coordinates
\((z_\cosm,\bar{z}_\cosm)\) produces a state that is smooth on the cosmological
horizon, but is singular at the black hole outer horizon except when
\(\kappa_+=\kappa_\cosm\). Since we are more interested in black holes with
smooth outer horizons, we instead opt for a compromise: the Unruh state with
outgoing Hawking radiation at temperature \(\kappa_+/2\pi\) and infalling
cosmological radiation at temperature \(\kappa_\cosm/2\pi\). In our CFT context,
this Unruh state is the conformal vacuum associated to the
\((z_+,\bar{z}_\cosm)\) coordinates. In the Lorentzian geometry, this Unruh
state will be smooth in the spacetime regions covered by this pair of Kruskal
coordinates. This is similar to the physically relevant parts of the spacetime
for a black hole formed by collapse, where the matter covers up the left
asymptotic region. This is one reason why the dS Unruh vacuum has been studied,
\eg{} by \cite{Hollands:2019whz}, for the purposes of probing SCC.

In principle, one can also place such a state on asymptotically flat black
holes, where there is a shower of radiation at some temperature
\(\kappa_c/2\pi\) supplied by the right past null infinity \(\mathscr{I}^-\).
Here, \(\kappa_\cosm\) is just a free parameter, not determined by geometry
\labelcref{eq:blackening_nearhorizon} as in the dS case.

\paragraph{Unruh state with no infalling radiation.}
However, perhaps more realistic in this case would be a state where there is no
such shower of radiation coming from \(\mathscr{I}^-\). This is the most natural
Unruh state for asymptotically flat black holes \cite{Unruh:1976db}. Since
\(\bar{z}_\tor\) approaches the Minkowski null coordinate \(r+t\) on
\(\mathscr{I}^-\), this Unruh state is the conformal vacuum associated to
\((z_+,\bar{z}_\tor)\). Again, the state is smooth on the Lorentzian spacetime
regions covered by these null coordinates.

\subsubsection{Energy and entanglement}
\label{sec:energyentanglement}

In each of the conformal vacua identified above, we will now calculate the
stress tensor expectation value near the Cauchy, inner horizon and the mutual
information \(I_{R_1:R_3}\) between the regions illustrated in
\cref{fig:stationarylor}. (The limit where the black entangling surface is
placed on the red inner horizon reproduces the regions \(R_1\) and \(R_3\)
discussed in \cref{sec:cauchyhorizons}.) We will first describe the general
calculation of these quantities before presenting the results specific to each
state.

To obtain the stress tensor expectation value, we first observe in Euclidean
signature that, on the complex plane and the disk with metric
\(\dd{z}\dd{\bar{z}}\), the expectation value vanishes. Here, we are bearing in
mind some holomorphic and anti-holomorphic coordinates \((z,\bar{z})\)
corresponding to the choice of conformal vacuum, as previously described.
Continuing to Lorentzian signature, where \((z,\bar{z})\) become null
coordinates, and reinstating the factor \(\Omega^{-2}\) in
\cref{eq:statmetric_weyl} by a Weyl transformation, the stress tensor
becomes\footnote{See, \eg{} eqs.~(6.136) and (6.137) in \cite{Birrell:1982ix}.}
\begin{align}
  \ev{T_{\bar{z}\bar{z}}}
  &= -\frac{c}{12\pi}\, \Omega^{-1} \partial_{\bar{z}}^2 \Omega
    \;,
    \label{eq:stressweyl}
\end{align}
with a similar \(T_{zz}\) component and, of course, a state-independent trace
determined by the curvature \labelcref{eq:statricci}. For relevance to SCC, we
are particularly interested in the near-inner-horizon divergences of the stress
tensor, which come from the above left-moving component as one approaches the
right inner horizon in \cref{fig:stationarylor}. More precisely, to separate
physical divergences from coordinate pathologies, one should transform
\cref{eq:stressweyl} to a coordinate basis that is regular at the inner horizon,
\eg{} \((z_-,\bar{z}_-)\). (Let us emphasize that this last step is just a
coordinate transformation which leaves the stress \emph{tensor} and metric
\emph{tensor} invariant, not a conformal transformation.)

To evaluate the mutual information \(I_{R_1:R_3}\), we need the entropies
\(S_{R_1}\), \(S_{R_3}\), and \(S_{R_1\cup R_3}\). We will write the coordinates
of the left endpoint of \(R_1\), the entangling surface separating \(R_1\) from
\(R_3\), and the right endpoint of \(R_3\) respectively as \((z_1,\bar{z}_1)\),
\((z_\ent,\bar{z}_\ent)\), and \((z_3,\bar{z}_3)\), where again we have in mind
some choice of null coordinates associated to a given conformal vacuum. We will
eventually send \((z_3,\bar{z}_3)\) to the past boundary of the right inner
horizon, as illustrated in \cref{fig:stationarylor}.

For asymptotically flat and dS black holes, the conformal vacua are prepared on
the full complex plane, where the two-point function of twist operators is
determined by conformal invariance. After analytic continuation to Lorentzian
signature and a Weyl transformation, the resulting entropies are
\begin{align}
  S_{R_1}
  &= S^{\pln}(
    z_1, \bar{z}_1;
    z_\ent, \bar{z}_\ent
    )
    - \frac{c}{6} \log\left[
    \Omega(z_1,\bar{z}_1)
    \Omega(z_\ent,\bar{z}_\ent)
    \right]
    \;,
    \label{eq:bhcft_entropyr1}
  \\
  S_{R_3}
  &= S^{\pln}(
    z_\ent, \bar{z}_\ent;
    z_3, \bar{z}_3
    )
    - \frac{c}{6} \log\left[
    \Omega(z_\ent,\bar{z}_\ent)
    \Omega(z_3,\bar{z}_3)
    \right]
    \;,
    \label{eq:bhcft_entropyr3}
  \\
  S_{R_1\cup R_3}
  &= S^{\pln}(
    z_1, \bar{z}_1;
    z_3, \bar{z}_3
    )
    - \frac{c}{6} \log\left[
    \Omega(z_1,\bar{z}_1)
    \Omega(z_3,\bar{z}_3)
    \right]
    \;,
    \label{eq:bhcft_entropyr1r3}
\end{align}
where \(S^\pln\) is again given by \cref{eq:entropy_twopoint}.

For asymptotically AdS black holes, the Hartle-Hawking state is prepared on the
Euclidean \(z\) disk which can be mapped to the \(w\) half-plane --- this is
illustrated in \cref{fig:stationaryeuc} and described by \cref{eq:disktoplane}.
As previously confronted in \cref{sec:joiningquench}, two-point functions in the
half-plane are not conformally determined. However, the limit
\((z_{3,+},\bar{z}_{3,+})\to(0,+\infty)\) where the right endpoint of \(R_3\) is
sent to the past boundary of the right inner horizon, as pictured in
\cref{fig:stationarylor}, corresponds to sending a twist operator in the \(w\)
half-plane much closer to the boundary than to other insertions.\footnote{In
  terms of the cross-ratio \labelcref{eq:cratio}, this corresponds to taking
  \(\eta(w(z_{\ent,+}),\bar{w}(\bar{z}_{\ent,+});w(z_{3,+}),\bar{w}(\bar{z}_{3,+}))\to0\)
  and
  \(\eta(w(z_{1,+}),\bar{w}(\bar{z}_{1,+});w(z_{3,+}),\bar{w}(\bar{z}_{3,+}))\to0\).
  This boundary limit is the opposite of the OPE limit described in
  \cref{foot:cratio_ope}. \label{foot:cratio_bdy}} Thus, the half-plane
two-point function is well-approximated by one-point functions. Consequently,
\begin{align}
  \begin{split}
    S_{R_3}
    &\sim
      S^{\uhp}(w(z_{\ent,+}), \bar{w}(\bar{z}_{\ent,+}))
      + S^{\uhp}(w(z_{3,+}), \bar{w}(\bar{z}_{3,+}))
    \\
    &\phantom{{}={}}
      - \frac{c}{6} \log\left[
      \Omega(z_{\ent,+},\bar{z}_{\ent,+})
      \Omega(z_{3,+},\bar{z}_{3,+})
      \sqrt{
      w'(z_{\ent,+})\bar{w}'(\bar{z}_{\ent,+})
      w'(z_{3,+})\bar{w}'(\bar{z}_{3,+})
      }
      \right]
      \;,
  \end{split}
  \label{eq:ads_entropyr3}
  \\
  \begin{split}
    S_{R_1\cup R_3}
    &\sim
      S^{\uhp}(w(z_{1,+}), \bar{w}(\bar{z}_{1,+}))
      + S^{\uhp}(w(z_{3,+}), \bar{w}(\bar{z}_{3,+}))
    \\
    &\phantom{{}={}}
      - \frac{c}{6} \log\left[
      \Omega(z_{1,+},\bar{z}_{1,+})
      \Omega(z_{3,+},\bar{z}_{3,+})
      \sqrt{
      w'(z_{1,+})\bar{w}'(\bar{z}_{1,+})
      w'(z_{3,+})\bar{w}'(\bar{z}_{3,+})
      }
      \right]
      \;,
  \end{split}
  \label{eq:ads_entropyr1r3}
\end{align}
where \(S^{\uhp}\) is given by \cref{eq:entropy_onepoint}. To obtain a similarly
universal answer for \(S_{R_1}\), we also take
\((z_{1,+},\bar{z}_{1,+})\to(-\infty,0)\) to the past boundary of the left inner
horizon, in which case
\begin{align}
  \begin{split}
    S_{R_1}
    &\sim
      S^{\uhp}(w(z_{1,+}), \bar{w}(\bar{z}_{1,+}))
      + S^{\uhp}(w(z_{\ent,+}), \bar{w}(\bar{z}_{\ent,+}))
    \\
    &\phantom{{}={}}
      - \frac{c}{6} \log\left[
      \Omega(z_{1,+},\bar{z}_{1,+})
      \Omega(z_{\ent,+},\bar{z}_{\ent,+})
      \sqrt{
      w'(z_{1,+})\bar{w}'(\bar{z}_{1,+})
      w'(z_{\ent,+})\bar{w}'(\bar{z}_{\ent,+})
      }
      \right]
      \;.
  \end{split}
  \label{eq:ads_entropyr1}
\end{align}

We will now explicitly evaluate the stress tensor near the inner horizon and the
mutual information \(I_{R_1:R_3}\) in each of the conformal vacua introduced in
\cref{sec:states}. The presentation will be in the reverse order in which the
states were introduced, to make it clear that the entanglement near the inner
horizon of an asymptotically AdS black hole is an outlier case, as expected from
the arguments of \cref{sec:cauchyhorizons}.

\paragraph{Unruh state with no infalling radiation.}
In the Unruh state with no infalling radiation, the stress tensor is given by
\cref{eq:stressweyl}, where the Weyl factor is given by
\cref{eq:weyl_flatunruh}. More precisely, \cref{eq:stressweyl} gives the
\(\ev{T_{\bar{z}_\tor \bar{z}_\tor}}\) component of the stress tensor.
Performing a coordinate transformation (not a conformal transformation, which
also rescales the metric tensor) to \((z_-,\bar{z}_-)\) and expanding near the
inner horizon, we find\footnote{Somewhat surprisingly, we have
  \(\order{(z_-\bar{z}_-)^2}\) instead of \(\order{z_-\bar{z}_-}\) for the
  relative size of the correction in \cref{eq:flatunruh_stressenergy}. The
  assumed expansion \labelcref{eq:blackening_nearhorizon_extra} of \(f\) noted
  in footnote \cref{foot:blackening_nearhorizon_extra} results in an exact
  cancellation of the naive next-to-leading correction in
  \cref{eq:flatunruh_stressenergy} and also
  \cref{eq:dsunruh_stressenergy,eq:hh_stressenergy} further below. Note also
  that \cref{eq:stressweyl} describes only conformal \emph{vacua}.}
\begin{align}
  \ev{T_{\bar{z}_- \bar{z}_-}}
  &= -\frac{c}{48 \pi \bar{z}_-^2}
    [1 + \order{(z_-\bar{z}_-)^2}]
    \;.
    \label{eq:flatunruh_stressenergy}
\end{align}

Combining \cref{eq:bhcft_entropyr1,eq:bhcft_entropyr3,eq:bhcft_entropyr1r3} with
\cref{eq:mi}, we find the mutual information
\begin{align}
  I_{R_1:R_3}
  &= \frac{c}{6} \log \left[
    \frac{z_{\ent,+}}{z_{1,+}}
    \frac{
    (z_{1,+} - z_{\ent,+})
    (\bar{z}_{1,\tor} -\bar{z}_{\ent,\tor})
    }{\Omega_{+\, \tor}^2(z_\ent,\bar{z}_\ent)}
    \right]
    \;,
\end{align}
where \(\Omega_{+\,\tor}\) is given by \cref{eq:weyl_flatunruh} and we have sent
the right endpoint of \(R_3\) to the past boundary of the right inner horizon
\((z_{3,+},\bar{z}_{3,\tor})\to(0,+\infty)\), as indicated in figure
\cref{fig:stationarylor}. Rewriting in terms of inner Kruskal coordinates and
pushing the entangling surface \((z_\ent, \bar{z}_\ent)\) towards the right
inner horizon by taking small \(-\bar{z}_{\ent,-}\), we find
\begin{align}
  I_{R_1:R_3}
  &= \frac{c}{6}
    \log\left[
    \bar{z}_{\ent,-}
    \log\left( \frac{\bar{z}_{\ent,-}}{\bar{z}_{1,-}} \right)
    \right]
    + (\text{constant in }\bar{z}_{\ent})
    + \order{z_{\ent,-}\bar{z}_{\ent,-}}
    \;.
    \label{eq:flatunruh_mi}
\end{align}

From \cref{eq:flatunruh_stressenergy,eq:flatunruh_mi}, we see that the stress
tensor has a negative quadratic divergence while the mutual information has a
negative logarithmic divergence in the limit \(\bar{z}_{\ent,-}\to 0^-\) where
the inner horizon is approached from below. As a sanity check, let us consider
the QNEC. Near the inner horizon, the Kruskal coordinate \(\bar{z}_- \sim
\lambda\) serves approximately, at leading order, as an affine parameter on the
blue null line along which the black entangling surface is pushed in
\cref{fig:stationarylor}. Thus, the QNEC states
\begin{align}
  \ev{T_{\bar{z}_- \bar{z}_-}(z_e, \bar{z}_e)}
  \sim \ev{T_{kk}(z_e, \bar{z}_e)}
  \ge \frac{1}{4\pi} \partial_\lambda^2 I_{R_1:R_3}
  \sim \frac{1}{4\pi} \partial_{z_{\ent,-}}^2 I_{R_1:R_3}
  \label{eq:statbh_qnec}
\end{align}
where \(k = \partial_\lambda \sim \partial_{\bar{z}_-}\). The leading
divergences of \cref{eq:flatunruh_stressenergy,eq:flatunruh_mi} are certainly
consistent with the QNEC.

By explicit calculation, we have shown that \(I_{R_1:R_3}\) diverges negatively
in the limit \(\bar{z}_{\ent,-}\to 0^-\) where \(R_3\) coincides with a past
portion the Cauchy, inner horizon, as predicted in \cref{sec:cauchyhorizons}. As
explained there, the mutual information \(I_{R_1:R_2}\) between the grey and
yellow subregions in \cref{fig:cauchyhorizon_flat} should also be negatively
divergent across the Cauchy horizon. By our argument in \cref{sec:scc}, SCC
should therefore be upheld (had we been considering a gravitational theory).

Indeed, we see that the stress tensor \labelcref{eq:flatunruh_stressenergy}
features a non-integrable divergence as the Cauchy horizon is approached ---
this is the signature of SCC that more standard approaches (such as
\cite{Hollands:2019whz}) usually look for. However, it should be noted that this
is not the infinite \emph{positive} energy promised by our analysis in
\cref{sec:disentanglement,sec:scc} as a consequence of disentanglement. As
already seen in the example considered in \cref{sec:joiningquench}, the infinite
positive energy which violates statement \labelcref{item:finiteenergy} appears
on (or immediately after) the surface --- here, the Cauchy horizon --- where the
disentanglement criterion \labelcref{item:disentanglement1} is satisfied, not on
approach to it. This infinite positive energy will only become apparent when one
attempts any (sensible, in the sense of statement
\labelcref{item:smoothnessafter1}) extension of the state beyond the Cauchy
horizon.

Nonetheless, we see that the QNEC \labelcref{eq:statbh_qnec} tells us that
whenever one finds negative, non-integrably divergent null energy on approach to
the Cauchy horizon, our disentanglement criterion
\labelcref{item:disentanglement1} and thus our SCC argument in \cref{sec:scc}
should also apply. Below, we will also see cases where the Cauchy horizon
satisfies our disentanglement criteria
\labelcref{item:disentanglement1,item:disentanglement3} even when the stress
tensor diverges positively or not at all as the Cauchy horizon is approached
from below.

\paragraph{Unruh state with infalling cosmological radiation.}
Next, let us consider the Unruh state with infalling radiation at temperature
\(\kappa_\cosm/2\pi\). In this case, the stress tensor
\(\ev{T_{\bar{z}_\cosm\bar{z}_\cosm}}\) is given by \cref{eq:stressweyl} with
the Weyl factor \labelcref{eq:weyl_dsunruh}. Written using inner horizon Kruskal
coordinates, we have
\begin{align}
  \ev{T_{\bar{z}_- \bar{z}_-}}
  &= \frac{c}{48 \pi}
    \frac{\kappa_\cosm^2 - \kappa_-^2}{\kappa_-^2}
    \frac{1}{\bar{z}_-^2}
    [1 + \order{(z_-\bar{z}_-)^2}]
    \;,
    \label{eq:dsunruh_stressenergy}
\end{align}
in agreement with the result of \cite{Hollands:2019whz}. The mutual information
between regions \(R_1\) and \(R_3\) is again obtained by combining
\cref{eq:bhcft_entropyr1,eq:bhcft_entropyr3,eq:bhcft_entropyr1r3}. Placing the
right endpoint \((z_{3,+},\bar{z}_{3,\cosm})\to(0,0)\) of \(R_3\) on the past
boundary the right inner horizon, the mutual information reads
\begin{align}
  I_{R_1:R_3}
  &= \frac{c}{6} \log \left[
    \frac{
    z_{\ent,+} \bar{z}_{\ent,\cosm}
    }{
    z_{1,+} \bar{z}_{1,\cosm}
    }
    \frac{
    (z_{1,+} - z_{\ent,+})(\bar{z}_{1,\cosm} - \bar{z}_{\ent,\cosm})
    }{\Omega_{+\, \cosm}^2(z_\ent,\bar{z}_\ent)}
    \right]
  \\
  &= \frac{c}{6}
    \log\left\{
    (-\bar{z}_{\ent,-})
    \left[
    (-\bar{z}_{1,-})^{\kappa_\cosm/\kappa_-}
    - (-\bar{z}_{\ent,-})^{\kappa_\cosm/\kappa_-}
    \right]
    \right\}
    + (\text{constant in }\bar{z}_{\ent})
    + \order{z_{\ent,-}\bar{z}_{\ent,-}}
    \;.
    \label{eq:dsunruh_mi}
\end{align}
In the last line, we have isolated the dependence on the small separation
\(-\bar{z}_{\ent,-}\) between the inner horizon and the entangling surface.
Again, the QNEC \labelcref{eq:statbh_qnec} is satisfied by the leading
divergences of \cref{eq:dsunruh_stressenergy,eq:dsunruh_mi}.

\Cref{eq:dsunruh_stressenergy} describes the competition between the negative
null energy background \labelcref{eq:flatunruh_stressenergy} previously seen in
the absence of infalling radiation, and the positive null energy of infalling
cosmological radiation. Depending on the temperatures of the inner horizon and
the cosmological radiation, the stress tensor can diverge positively,
negatively, or not at all as one approaches the inner horizon \(\bar{z}_- \to
0^-\).

In contrast, as \(\bar{z}_{\ent,-}\to 0^-\), the negative divergence of the
mutual information \labelcref{eq:dsunruh_mi} is insensitive to these
temperatures. As explained in \cref{sec:cauchyhorizons}, the dS\footnote{As
  previously described, we can also consider this same conformal vacuum on
  asymptotically flat backgrounds, with infalling ``cosmological'' radiation at
  temperature \(\kappa_\cosm/2\pi\) not necessarily dictated by geometry. This
  is the same as the Hartle-Hawking state when \(\kappa_\cosm=\kappa_+\).
  Otherwise, the state is singular on the \(\bar{z}_+=0\) piece of the outer
  horizon. Regardless, the interpretation of the negative divergence of mutual
  information is the same as described for the Hartle-Hawking state below.}
cosmological radiation falling into the black hole is entangled with partners
across the cosmological horizon that are captured by the dS future infinity
\(\mathscr{I}^+\). Such radiation does not therefore contribute to
\(I_{R_1:R_3}\) significantly to alter its divergent behaviour as \(R_3\)
approaches a past portion of the inner, Cauchy horizon. Our argument for SCC in
\cref{sec:scc,sec:cauchyhorizons} based on this negative divergence in mutual
information therefore always applies, even when more standard
stress-tensor-based approaches would see no divergence at \(\kappa_\cosm =
\kappa_-\).

\paragraph{Hartle-Hawking state.}
Finally, we consider the Hartle-Hawking state. Here, \cref{eq:stressweyl} gives
\(\ev{T_{\bar{z}_+ \bar{z}_+}}\) in terms of the Weyl factor
\labelcref{eq:weyl_hh}. In inner Kruskal coordinates, this becomes
\begin{align}
  \ev{T_{\bar{z}_- \bar{z}_-}}
  &= \frac{c}{48 \pi}
    \frac{\kappa_+^2 - \kappa_-^2}{\kappa_-^2}
    \frac{1}{\bar{z}_-^2}
    [1 + \order{(z_-\bar{z}_-)^2}]
    \;.
    \label{eq:hh_stressenergy}
\end{align}
It has an interpretation similar to \cref{eq:dsunruh_stressenergy}; in lieu of
cosmological radiation, the black hole now absorbs radiation at the outer
horizon temperature \(\kappa_+/2\pi\). The Hartle-Hawking state is equally
applicable to asymptotically flat and AdS black holes, which both share the
above near-inner-horizon behaviour of the stress tensor.\footnote{In principle,
  one can technically consider asymptotically dS black holes in the
  Hartle-Hawking state, subject to the caveats described below
  \cref{eq:disktoplane}. Proceeding anyway,
  \cref{eq:hh_stressenergy,eq:dsflathh_mi_exact,eq:dsflathh_mi_approx} would be
  applicable in this case.}

There is a drastic difference between the two cases, however, in the mutual
information \(I_{R_1:R_3}\). Let us consider asymptotically flat black holes
first. \Cref{eq:bhcft_entropyr1,eq:bhcft_entropyr3,eq:bhcft_entropyr1r3} are
applicable in this case. Combining these into a mutual information
\labelcref{eq:mi} and, as always, placing the right endpoint
\((z_{3,+},\bar{z}_{3,+})\to(0,+\infty)\) of \(R_3\) on the past boundary of the
right inner horizon, we obtain
\begin{align}
  I_{R_1:R_3}
  &= \frac{c}{6} \log \left[
    \frac{z_{\ent,+}}{z_{1,+}}
    \frac{
    (z_{1,+} - z_{\ent,+})
    (\bar{z}_{1,+} -\bar{z}_{\ent,+})
    }{\Omega_{+ +}^2(z_\ent,\bar{z}_\ent)}
    \right]
    \label{eq:dsflathh_mi_exact}
  \\
  &= \frac{c}{6}
    \log\left\{
    (-\bar{z}_{\ent,-})
    \left[
    (-\bar{z}_{1,-})^{\kappa_+/\kappa_-}
    - (-\bar{z}_{\ent,-})^{\kappa_+/\kappa_-}
    \right]
    \right\}
    + (\text{constant in }\bar{z}_{\ent})
    + \order{z_{\ent,-}\bar{z}_{\ent,-}}
    \;.
    \label{eq:dsflathh_mi_approx}
\end{align}
The behaviour \labelcref{eq:dsflathh_mi_approx}, as the entangling surface
\(\bar{z}_{\ent,-}\to 0^-\) is pushed towards the inner horizon, is similar to
\cref{eq:dsunruh_mi}. Again, the only difference is that the temperature of the
infalling radiation has been changed from \(\kappa_\cosm/2\pi\) to
\(\kappa_+/2\pi\). Because the Hartle-Hawking state is a thermofield double, the
radiation falling in from the right exterior, in particular the \(r_1\)
discussed in \cref{sec:cauchyhorizons}, is entangled primarily with outgoing
radiation in the left exterior captured by \(\mathscr{I}^+\). Consequently, this
radiation, much like dS cosmological radiation, does not alter the divergent
behaviour of the above mutual information in the \(\bar{z}_{\ent,-}\to 0^-\)
limit.

The AdS asymptotic boundary \(\mathscr{I}\), however, reflects outgoing
radiation back into the black hole. Consequently, we concluded in
\cref{sec:cauchyhorizons} that \(I_{R_1:R_3}\) need not be negatively divergent
in the limit \(\bar{z}_{\ent,-} \to 0^-\) where \(R_3\) coincides with a past
portion of the inner, Cauchy horizon. Let us now see this explicitly from the
entropy formulas
\labelcref{eq:ads_entropyr1,eq:ads_entropyr3,eq:ads_entropyr1r3} derived in the
presence of the boundary. These combine to give the mutual information
\begin{align}
  I_{R_1:R_3}
  &= 2 S^{\uhp}(w(z_{\ent,+}), \bar{w}(\bar{z}_{\ent,+}))
    - \frac{c}{3} \log\left[
    \Omega(z_{\ent,+},\bar{z}_{\ent,+})
    \sqrt{
    w'(z_{\ent,+})\bar{w}'(\bar{z}_{\ent,+})
    }
    \right]
    \label{eq:adshh_mi_exact}
  \\
  &= \frac{c}{6} \log\left\{
    - \frac{
    (-\bar{z}_{\ent,-})^{1+\frac{\kappa_+}{\kappa_-}}
    \left[
    w(z_{\ent,+}) - \bar{w}(\bar{z}_{\ent,+})
    \right]^2
    }{
    \bar{w}'(\bar{z}_{\ent,+})
    }
    \right\}
    + (\text{constant in }\bar{z}_{\ent})
    + \order{z_{\ent,-}\bar{z}_{\ent,-}}
    \;.
    \label{eq:adshh_mi_approx}
\end{align}
In \cref{eq:adshh_mi_exact}, the left endpoint of \(R_1\) and the right endpoint
of \(R_3\) have been placed respectively at the past boundaries of the left and
right inner horizons.\footnote{Notice here that, relative to cases we previously
  considered, we have fixed an extra freedom by pushing the left endpoint of
  \(R_1\) to the past boundary of the left inner horizon. As explained above
  \cref{eq:ads_entropyr1} and in \cref{foot:cratio_bdy}, we did this to evaluate
  a two-point function in the half-plane by taking an extreme limit
  \(\eta(w(z_{1,+}),\bar{w}(\bar{z}_{1,+});w(z_{\ent,+}),\bar{w}(\bar{z}_{\ent,+}))\to0\)
  of the cross-ratio defined in \cref{eq:cratio}.

  One might wonder if this is the underlying reason why the
  \(\bar{z}_{\ent,-}\to 0^-\) divergence we are finding for \(I_{R_1:R_3}\) in
  AdS is so different from previously considered cases --- let us dissuade
  against this thinking. Suppose that \(\eta\) had not been taken to an extreme
  value, \(0\) or \(\infty\), by our choice of the left endpoint of \(R_1\).
  Then the first line of \cref{eq:ads_entropyr1} would generally be replaced by
  something with also nontrivial dependence on \(\eta\), not fixed by conformal
  invariance. However, just as a half-plane two-point function should be finite
  except when the insertions collide with each other or with the boundary, the
  corrected first line of \cref{eq:ads_entropyr1} should remain finite apart
  from \(\eta\to0,\infty\). Taking \(\bar{z}_{\ent,-}\to 0^-\) at fixed
  \(z_{\ent,+}<0\) does not send \(\eta\to0,\infty\) unless \(\eta\) was already
  at these extreme values to begin with. Thus, whatever nontrivial
  \(\eta\)-dependence acquired by the corrected first line of
  \cref{eq:ads_entropyr1} cannot contribute to divergences in \(I_{R_1:R_3}\) as
  \(\bar{z}_{\ent,-}\to 0^-\). The only possible divergences are from the
  second, transformation term in \cref{eq:adshh_mi_exact}, just as we find
  below.} In fact, this can be viewed as removing the left endpoint of \(R_1\)
and the right endpoint of \(R_3\) so that \(R_1\) and \(R_3\) become
complementary to each other. Thus, the above is equal to double the entropy of
\(R_1\) with no left endpoint or, equivalently, double the entropy of \(R_3\)
with no right endpoint.

For small \(\bar{z}_{\ent,-}\), note that
\begin{align}
  \bar{w}(\bar{z}_{\ent,+})
  &= i + \frac{2 R_{\mathscr{I}}}{\bar{z}_{\ent,+}}
    + \order{\bar{z}_{\ent,+}^{-2}}
    = i + 2 R_{\mathscr{I}} (-\bar{z}_{\ent,-})^{\kappa_+/\kappa_-}
    + \order{(-\bar{z}_{\ent,-})^{2 \kappa_+/\kappa_-}}
    \;,
\end{align}
so the \([w(z_{\ent,+}) - \bar{w}(\bar{z}_{\ent,+})]^2\) factor in the logarithm
of \cref{eq:adshh_mi_approx} is innocuous. It is the
\((-\bar{z}_{\ent,-})^{1+\frac{\kappa_+}{\kappa_-}}\) factor together with
\begin{align}
  \bar{w}'(\bar{z}_{\ent,+})
  &= - \frac{2 R_{\mathscr{I}}}{\bar{z}_{\ent,+}^2}
    + \order{\bar{z}_{\ent,+}^{-3}}
    = - 2 R_{\mathscr{I}} (-\bar{z}_{\ent,-})^{2\kappa_+/\kappa_-}
    + \order{(-\bar{z}_{\ent,-})^{3\kappa_+/\kappa_-}}
\end{align}
which determines the divergent behaviour of \cref{eq:adshh_mi_approx} as
\(\bar{z}_{\ent,-}\to 0^-\). These factors originate from the second term in
\cref{eq:adshh_mi_exact} describing the conformal and Weyl transformation
between the \(w\) half-plane and the black hole geometry. We see that the
leading divergence
\(\frac{c}{6}\frac{\kappa_+-\kappa_-}{\kappa_-}\log(-1/\bar{z}_{\ent,-})\) of
\cref{eq:adshh_mi_approx} is consistent with the QNEC \labelcref{eq:statbh_qnec}
and the stress tensor \labelcref{eq:hh_stressenergy}.

This divergence in mutual information can be negative, zero, or positive,
depending on whether the outer horizon temperature \(\kappa_+/2\pi\) is less
than, equal to, or greater than the inner horizon temperature \(\kappa_-/2\pi\),
in contrast to all cases we have considered previously. Of course, this is
consistent with our discussion in \cref{sec:cauchyhorizons}, where we explained
how Cauchy horizons in AdS black holes may present outlying exceptions to our
disentanglement criteria. From our disentanglement argument alone, it may
therefore appear that mutual information is a weaker probe of SCC in AdS than
the stress tensor, which diverges except when \(\kappa_+=\kappa_-\). However,
let us point out that positive divergences of mutual information are also
dangerous: instead of applying our disentanglement argument, one can appeal
directly to the QNEC \labelcref{eq:statbh_qnec}, which states \eg{} that a
positive logarithmic divergence in mutual information (or entropy) leads to a
positive quadratic divergence in the stress tensor. The QFC
\labelcref{eq:qfc_mi} is similarly helpful, with dynamical
gravity.\footnote{Considering \(I_{R_1:R_3}\), one might worry about the
  validity of the QFC due to the large extrinsic curvature \cite{Bousso:2022tdb}
  present in the slice containing \(R_1\cup R_3\) as the entangling surface is
  pushed to the Cauchy, inner horizon. But, as mentioned below
  \cref{eq:adshh_mi_approx}, for AdS, it suffices to work with just the entropy
  of \(R_1\).}

\bibliographystyle{JHEP.bst}
\bibliography{arxiv_v2.bib}

\providecommand{\href}[2]{#2}\begingroup\raggedright\begin{thebibliography}{10}

\bibitem{VanRaamsdonk:2010pw}
M.~Van~Raamsdonk, \emph{{Building up spacetime with quantum entanglement}},
  \href{https://doi.org/10.1142/S0218271810018529}{\emph{Gen. Rel. Grav.}
  {\bfseries 42} (2010) 2323}
  [\href{https://arxiv.org/abs/1005.3035}{{\ttfamily 1005.3035}}].

\bibitem{Maldacena:2013xja}
J.~Maldacena and L.~Susskind, \emph{{Cool horizons for entangled black holes}},
  \href{https://doi.org/10.1002/prop.201300020}{\emph{Fortsch. Phys.}
  {\bfseries 61} (2013) 781} [\href{https://arxiv.org/abs/1306.0533}{{\ttfamily
  1306.0533}}].

\bibitem{Almheiri:2019psf}
A.~Almheiri, N.~Engelhardt, D.~Marolf and H.~Maxfield, \emph{{The entropy of
  bulk quantum fields and the entanglement wedge of an evaporating black
  hole}}, \href{https://doi.org/10.1007/JHEP12(2019)063}{\emph{JHEP} {\bfseries
  12} (2019) 063} [\href{https://arxiv.org/abs/1905.08762}{{\ttfamily
  1905.08762}}].

\bibitem{Penington:2019npb}
G.~Penington, \emph{{Entanglement Wedge Reconstruction and the Information
  Paradox}}, \href{https://doi.org/10.1007/JHEP09(2020)002}{\emph{JHEP}
  {\bfseries 09} (2020) 002}
  [\href{https://arxiv.org/abs/1905.08255}{{\ttfamily 1905.08255}}].

\bibitem{Hawking:1976ra}
S.W.~Hawking, \emph{{Breakdown of Predictability in Gravitational Collapse}},
  \href{https://doi.org/10.1103/PhysRevD.14.2460}{\emph{Phys. Rev. D}
  {\bfseries 14} (1976) 2460}.

\bibitem{Almheiri:2012rt}
A.~Almheiri, D.~Marolf, J.~Polchinski and J.~Sully, \emph{{Black Holes:
  Complementarity or Firewalls?}},
  \href{https://doi.org/10.1007/JHEP02(2013)062}{\emph{JHEP} {\bfseries 02}
  (2013) 062} [\href{https://arxiv.org/abs/1207.3123}{{\ttfamily 1207.3123}}].

\bibitem{Czech:2012be}
B.~Czech, J.L.~Karczmarek, F.~Nogueira and M.~Van~Raamsdonk, \emph{{Rindler
  Quantum Gravity}},
  \href{https://doi.org/10.1088/0264-9381/29/23/235025}{\emph{Class. Quant.
  Grav.} {\bfseries 29} (2012) 235025}
  [\href{https://arxiv.org/abs/1206.1323}{{\ttfamily 1206.1323}}].

\bibitem{Emparan:2023ypa}
R.~Emparan and J.M.~Magan, \emph{{Tearing down spacetime with quantum
  disentanglement}}, \href{https://doi.org/10.1007/JHEP03(2024)078}{\emph{JHEP}
  {\bfseries 03} (2024) 078}
  [\href{https://arxiv.org/abs/2312.06803}{{\ttfamily 2312.06803}}].

\bibitem{Bousso:2015mna}
R.~Bousso, Z.~Fisher, S.~Leichenauer and A.C.~Wall, \emph{{Quantum focusing
  conjecture}}, \href{https://doi.org/10.1103/PhysRevD.93.064044}{\emph{Phys.
  Rev. D} {\bfseries 93} (2016) 064044}
  [\href{https://arxiv.org/abs/1506.02669}{{\ttfamily 1506.02669}}].

\bibitem{Bousso:2015wca}
R.~Bousso, Z.~Fisher, J.~Koeller, S.~Leichenauer and A.C.~Wall, \emph{{Proof of
  the Quantum Null Energy Condition}},
  \href{https://doi.org/10.1103/PhysRevD.93.024017}{\emph{Phys. Rev. D}
  {\bfseries 93} (2016) 024017}
  [\href{https://arxiv.org/abs/1509.02542}{{\ttfamily 1509.02542}}].

\bibitem{Penrose:1974cup}
R.~Penrose, \emph{Gravitational collapse},
  \href{https://doi.org/10.1017/S007418090023605X}{\emph{Symposium -
  International Astronomical Union} {\bfseries 64} (1974) 82–91}.

\bibitem{Hollands:2019whz}
S.~Hollands, R.M.~Wald and J.~Zahn, \emph{{Quantum instability of the Cauchy
  horizon in Reissner\textendash{}Nordstr\"om\textendash{}deSitter spacetime}},
  \href{https://doi.org/10.1088/1361-6382/ab8052}{\emph{Class. Quant. Grav.}
  {\bfseries 37} (2020) 115009}
  [\href{https://arxiv.org/abs/1912.06047}{{\ttfamily 1912.06047}}].

\bibitem{Zilberman:2019buh}
N.~Zilberman, A.~Levi and A.~Ori, \emph{{Quantum fluxes at the inner horizon of
  a spherical charged black hole}},
  \href{https://doi.org/10.1103/PhysRevLett.124.171302}{\emph{Phys. Rev. Lett.}
  {\bfseries 124} (2020) 171302}
  [\href{https://arxiv.org/abs/1906.11303}{{\ttfamily 1906.11303}}].

\bibitem{Hollands:2020qpe}
S.~Hollands, C.~Klein and J.~Zahn, \emph{{Quantum stress tensor at the Cauchy
  horizon of the Reissner\textendash{}Nordstr\"om\textendash{}de Sitter
  spacetime}}, \href{https://doi.org/10.1103/PhysRevD.102.085004}{\emph{Phys.
  Rev. D} {\bfseries 102} (2020) 085004}
  [\href{https://arxiv.org/abs/2006.10991}{{\ttfamily 2006.10991}}].

\bibitem{Zilberman:2022aum}
N.~Zilberman, M.~Casals, A.~Ori and A.C.~Ottewill, \emph{{Quantum Fluxes at the
  Inner Horizon of a Spinning Black Hole}},
  \href{https://doi.org/10.1103/PhysRevLett.129.261102}{\emph{Phys. Rev. Lett.}
  {\bfseries 129} (2022) 261102}
  [\href{https://arxiv.org/abs/2203.08502}{{\ttfamily 2203.08502}}].

\bibitem{Papadodimas:2019msp}
K.~Papadodimas, S.~Raju and P.~Shrivastava, \emph{{A simple quantum test for
  smooth horizons}}, \href{https://doi.org/10.1007/JHEP12(2020)003}{\emph{JHEP}
  {\bfseries 12} (2020) 003}
  [\href{https://arxiv.org/abs/1910.02992}{{\ttfamily 1910.02992}}].

\bibitem{Unruh:1976db}
W.G.~Unruh, \emph{{Notes on black hole evaporation}},
  \href{https://doi.org/10.1103/PhysRevD.14.870}{\emph{Phys. Rev. D} {\bfseries
  14} (1976) 870}.

\bibitem{Susskind:1994sm}
L.~Susskind and J.~Uglum, \emph{{Black hole entropy in canonical quantum
  gravity and superstring theory}},
  \href{https://doi.org/10.1103/PhysRevD.50.2700}{\emph{Phys. Rev. D}
  {\bfseries 50} (1994) 2700}
  [\href{https://arxiv.org/abs/hep-th/9401070}{{\ttfamily hep-th/9401070}}].

\bibitem{Bousso:2015eda}
R.~Bousso and N.~Engelhardt, \emph{{Generalized Second Law for Cosmology}},
  \href{https://doi.org/10.1103/PhysRevD.93.024025}{\emph{Phys. Rev. D}
  {\bfseries 93} (2016) 024025}
  [\href{https://arxiv.org/abs/1510.02099}{{\ttfamily 1510.02099}}].

\bibitem{Almheiri:2019yqk}
A.~Almheiri, R.~Mahajan and J.~Maldacena, \emph{{Islands outside the horizon}},
   \href{https://arxiv.org/abs/1910.11077}{{\ttfamily 1910.11077}}.

\bibitem{Hartman:2020khs}
T.~Hartman, Y.~Jiang and E.~Shaghoulian, \emph{{Islands in cosmology}},
  \href{https://doi.org/10.1007/JHEP11(2020)111}{\emph{JHEP} {\bfseries 11}
  (2020) 111} [\href{https://arxiv.org/abs/2008.01022}{{\ttfamily
  2008.01022}}].

\bibitem{Akers:2016ugt}
C.~Akers, J.~Koeller, S.~Leichenauer and A.~Levine, \emph{{Geometric
  Constraints from Subregion Duality Beyond the Classical Regime}},
  \href{https://arxiv.org/abs/1610.08968}{{\ttfamily 1610.08968}}.

\bibitem{Akers:2017ttv}
C.~Akers, V.~Chandrasekaran, S.~Leichenauer, A.~Levine and
  A.~Shahbazi~Moghaddam, \emph{{Quantum null energy condition, entanglement
  wedge nesting, and quantum focusing}},
  \href{https://doi.org/10.1103/PhysRevD.101.025011}{\emph{Phys. Rev. D}
  {\bfseries 101} (2020) 025011}
  [\href{https://arxiv.org/abs/1706.04183}{{\ttfamily 1706.04183}}].

\bibitem{Wall:2011kb}
A.C.~Wall, \emph{{Testing the Generalized Second Law in 1+1 dimensional
  Conformal Vacua: An Argument for the Causal Horizon}},
  \href{https://doi.org/10.1103/PhysRevD.85.024015}{\emph{Phys. Rev. D}
  {\bfseries 85} (2012) 024015}
  [\href{https://arxiv.org/abs/1105.3520}{{\ttfamily 1105.3520}}].

\bibitem{Balakrishnan:2017bjg}
S.~Balakrishnan, T.~Faulkner, Z.U.~Khandker and H.~Wang, \emph{{A General Proof
  of the Quantum Null Energy Condition}},
  \href{https://doi.org/10.1007/JHEP09(2019)020}{\emph{JHEP} {\bfseries 09}
  (2019) 020} [\href{https://arxiv.org/abs/1706.09432}{{\ttfamily
  1706.09432}}].

\bibitem{Parentani:1993yz}
R.~Parentani, \emph{{The Energy momentum tensor in Fulling-Rindler vacuum}},
  \href{https://doi.org/10.1088/0264-9381/10/7/016}{\emph{Class. Quant. Grav.}
  {\bfseries 10} (1993) 1409}
  [\href{https://arxiv.org/abs/hep-th/9303062}{{\ttfamily hep-th/9303062}}].

\bibitem{Calabrese:2007mtj}
P.~Calabrese and J.~Cardy, \emph{{Entanglement and correlation functions
  following a local quench: a conformal field theory approach}},
  \href{https://doi.org/10.1088/1742-5468/2007/10/P10004}{\emph{J. Stat. Mech.}
  {\bfseries 0710} (2007) P10004}
  [\href{https://arxiv.org/abs/0708.3750}{{\ttfamily 0708.3750}}].

\bibitem{Ugajin:2013xxa}
T.~Ugajin, \emph{{Two dimensional quantum quenches and holography}},
  \href{https://arxiv.org/abs/1311.2562}{{\ttfamily 1311.2562}}.

\bibitem{Shimaji:2018czt}
T.~Shimaji, T.~Takayanagi and Z.~Wei, \emph{{Holographic Quantum Circuits from
  Splitting/Joining Local Quenches}},
  \href{https://doi.org/10.1007/JHEP03(2019)165}{\emph{JHEP} {\bfseries 03}
  (2019) 165} [\href{https://arxiv.org/abs/1812.01176}{{\ttfamily
  1812.01176}}].

\bibitem{Anderson:1986ww}
A.~Anderson and B.S.~DeWitt, \emph{{Does the Topology of Space Fluctuate?}},
  \href{https://doi.org/10.1007/BF01889374}{\emph{Found. Phys.} {\bfseries 16}
  (1986) 91}.

\bibitem{Penrose:1968ar}
R.~Penrose, \emph{{Structure of space-time}},  in \emph{{Battelle Rencontres}},
  pp.~121--235, 1968.

\bibitem{Hintz:2015jkj}
P.~Hintz and A.~Vasy, \emph{{Analysis of linear waves near the Cauchy horizon
  of cosmological black holes}},
  \href{https://doi.org/10.1063/1.4996575}{\emph{J. Math. Phys.} {\bfseries 58}
  (2017) 081509} [\href{https://arxiv.org/abs/1512.08004}{{\ttfamily
  1512.08004}}].

\bibitem{Cardoso:2017soq}
V.~Cardoso, J.a.L.~Costa, K.~Destounis, P.~Hintz and A.~Jansen,
  \emph{{Quasinormal modes and Strong Cosmic Censorship}},
  \href{https://doi.org/10.1103/PhysRevLett.120.031103}{\emph{Phys. Rev. Lett.}
  {\bfseries 120} (2018) 031103}
  [\href{https://arxiv.org/abs/1711.10502}{{\ttfamily 1711.10502}}].

\bibitem{Dias:2019ery}
O.J.C.~Dias, H.S.~Reall and J.E.~Santos, \emph{{The BTZ black hole violates
  strong cosmic censorship}},
  \href{https://doi.org/10.1007/JHEP12(2019)097}{\emph{JHEP} {\bfseries 12}
  (2019) 097} [\href{https://arxiv.org/abs/1906.08265}{{\ttfamily
  1906.08265}}].

\bibitem{Balasubramanian:2019qwk}
V.~Balasubramanian, A.~Kar and G.~S\'arosi, \emph{{Holographic Probes of Inner
  Horizons}}, \href{https://doi.org/10.1007/JHEP06(2020)054}{\emph{JHEP}
  {\bfseries 06} (2020) 054}
  [\href{https://arxiv.org/abs/1911.12413}{{\ttfamily 1911.12413}}].

\bibitem{Emparan:2020rnp}
R.~Emparan and M.~Toma\v{s}evi\'c, \emph{{Strong cosmic censorship in the BTZ
  black hole}}, \href{https://doi.org/10.1007/JHEP06(2020)038}{\emph{JHEP}
  {\bfseries 06} (2020) 038}
  [\href{https://arxiv.org/abs/2002.02083}{{\ttfamily 2002.02083}}].

\bibitem{Kolanowski:2023hvh}
M.~Kolanowski and M.~Toma\v{s}evi\'c, \emph{{Singularities in 2D and 3D quantum
  black holes}}, \href{https://doi.org/10.1007/JHEP12(2023)102}{\emph{JHEP}
  {\bfseries 12} (2023) 102}
  [\href{https://arxiv.org/abs/2310.06014}{{\ttfamily 2310.06014}}].

\bibitem{Shrivastava:2020xmw}
P.~Shrivastava, \emph{{Quantum aspects of charged black holes in de-Sitter
  space}},  \href{https://arxiv.org/abs/2009.03261}{{\ttfamily 2009.03261}}.

\bibitem{Chen:2019uhq}
H.Z.~Chen, Z.~Fisher, J.~Hernandez, R.C.~Myers and S.-M.~Ruan,
  \emph{{Information Flow in Black Hole Evaporation}},
  \href{https://doi.org/10.1007/JHEP03(2020)152}{\emph{JHEP} {\bfseries 03}
  (2020) 152} [\href{https://arxiv.org/abs/1911.03402}{{\ttfamily
  1911.03402}}].

\bibitem{Chen:2020jvn}
H.Z.~Chen, Z.~Fisher, J.~Hernandez, R.C.~Myers and S.-M.~Ruan,
  \emph{{Evaporating Black Holes Coupled to a Thermal Bath}},
  \href{https://doi.org/10.1007/JHEP01(2021)065}{\emph{JHEP} {\bfseries 01}
  (2021) 065} [\href{https://arxiv.org/abs/2007.11658}{{\ttfamily
  2007.11658}}].

\bibitem{Stanford:2022fdt}
D.~Stanford and Z.~Yang, \emph{{Firewalls from wormholes}},
  \href{https://arxiv.org/abs/2208.01625}{{\ttfamily 2208.01625}}.

\bibitem{Affleck:1991tk}
I.~Affleck and A.W.W.~Ludwig, \emph{{Universal noninteger 'ground state
  degeneracy' in critical quantum systems}},
  \href{https://doi.org/10.1103/PhysRevLett.67.161}{\emph{Phys. Rev. Lett.}
  {\bfseries 67} (1991) 161}.

\bibitem{Birrell:1982ix}
N.D.~Birrell and P.C.W.~Davies, \emph{{Quantum Fields in Curved Space}},
  Cambridge Monographs on Mathematical Physics, Cambridge Univ. Press,
  Cambridge, UK (2, 1984),
  \href{https://doi.org/10.1017/CBO9780511622632}{10.1017/CBO9780511622632}.

\bibitem{Bousso:2022tdb}
R.~Bousso and A.~Shahbazi-Moghaddam, \emph{{Quantum singularities}},
  \href{https://doi.org/10.1103/PhysRevD.107.066002}{\emph{Phys. Rev. D}
  {\bfseries 107} (2023) 066002}
  [\href{https://arxiv.org/abs/2206.07001}{{\ttfamily 2206.07001}}].

\end{thebibliography}\endgroup
\end{document}